%% file: dm-ddd-NR-arxiv.tex
\documentclass[12pt, fleqn, a4paper]{article} 
\usepackage{hyperref}
\hypersetup{pdfstartview = FitH,
            pdfborder    = {0 0 0.5},
            colorlinks   = false
}
 \usepackage{epsfig, subcaption}
 \usepackage{multirow}
 \usepackage{clshan-math, clshan-dm-dd, clshan-dm-ddd}
 \def \lsim {\:\raisebox{-0.7 ex}{$\stackrel{\textstyle<}{\sim}$}\:}
\input{def-newcommands}
\begin{document}
\thispagestyle{empty}
\begin{flushright}
 March 2021
\end{flushright}
\begin{center}
{\Large\bf
 Simulations of
 the Angular Recoil--Energy Distribution of     \\
 WIMP--Scattered Target Nuclei for              \\ \vspace{0.2 cm}
 Directional Dark Matter Detection Experiments} \\
\vspace*{0.7 cm}
 {\sc Chung-Lin Shan}                           \\
\vspace{0.5 cm}
 {\small\it
  Preparatory Office of
  the Supporting Center for
  Taiwan Independent Researchers                       \\ \vspace{0.05 cm}
  P.O.BOX 21 National Yang Ming Chiao Tung University,
  Hsinchu City 30099, Taiwan, R.O.C.}                  \\~\\~\\
 {\it E-mail:} {\tt clshan@tir.tw}
\end{center}
\vspace{2 cm}
\begin{abstract}

 In this paper,
 as the second part of the third step of
 our study on developing data analysis procedures
 for using 3-dimensional information
 offered by directional direct Dark Matter detection experiments
 in the future,
 we investigate
 the angular distributions of
 the recoil direction (flux)
 and the recoil energy of
 the Monte Carlo simulated WIMP--scattered target nuclei
 observed in different celestial coordinate systems.
 The ``anisotropy''
 and the ``directionality'' (``annual'' modulation) of
 the angular recoil--direction/energy distributions
 will be demonstrated.
 We will also discuss
 their dependences
 on the target nucleus
 and on the mass of incident halo WIMPs.
 For readers' reference,
 all simulation plots presented in this paper (and more)
 can be found ``in animation''
 on our online (interactive) demonstration webpage
 ({\tt \url{http://www.tir.tw/phys/hep/dm/amidas-2d/}}).

\end{abstract}
\clearpage
\section{Introduction}

 In the last (more than)
 three decades,
 a large number of experiments has been built
 and is being planned
 to search for the most favorite Dark Matter (DM) candidate:
 Weakly Interacting Massive Particles (WIMPs) $\chi$,
 by
 direct detection of
 the scattering recoil energy
 of ambient WIMPs off target nuclei
 in low--background underground laboratory detectors
 (see Refs.~%
  \cite{SUSYDM96,
        Gaitskell04,
        Baudis12c,
        Baudis20}
  for reviews).

 Besides non--directional direct detection experiments
 measuring only recoil energies
 deposited in detectors,
 the ``directional'' detection of Galactic DM particles
 has been proposed more than one decade
 to be a promising experimental strategy
 for discriminating signals from backgrounds
 by using additional 3-dimensional information
 (recoil tracks and/or head--tail senses)
 of (elastic) WIMP--nucleus scattering events
 (see Refs.~%
  \cite{Ahlen09,
        Mayet16, Battat16b,
        Vahsen20,
        Vahsen21}
  for reviews and recent progresses).

 As the preparation
 for our future study
 on the development of data analysis procedures
 for using and/or combining 3-D information
 offered by directional Dark Matter detection experiments
 to,
 e.g.,
 reconstruct the 3-dimensional WIMP velocity distribution,
 in Ref.~\cite{DMDDD-N},
 we started with the Monte Carlo generation of
 the 3-D velocity of
 (incident) halo WIMPs
 in the Galactic coordinate system,
 including
 the magnitude,
 the direction,
 and the incoming/scattering time.
 Each generated 3-D WIMP velocity
 has then been transformed to
 the laboratory--independent
 (Ecliptic,
  Equatorial,
  and Earth)
 coordinate systems
 as well as
 to the laboratory--dependent
 (horizontal and laboratory)
 coordinate systems
 (see the simulation workflow
  sketched in Fig.~\ref{fig:workflow})
 for the investigations on
 the angular distribution patterns of
 the 3-D WIMP velocity (flux)
 and the (accumulated and average) kinetic energy
 in different celestial coordinate systems
 \cite{DMDDD-N, DMDDD-P}
 as well as
 the Bayesian reconstruction of
 the radial component (magnitude) of
 the 3-D WIMP velocity
 \cite{DMDDD-N}.
 Not only
 the diurnal modulations,
 we demonstrated also
 the ``annual'' modulations of
 the angular WIMP velocity (flux)/kinetic--energy distributions
 \cite{DMDDD-N, DMDDD-P}.

 However,
 besides recoil energies,
 what one could measure (directly)
 in directional DM detection experiments
 is recoil tracks (with the sense--recognition)
 and in turn recoil angles (directions)
 of scattered target nuclei.
 In Ref.~\cite{DMDDD-3D-WIMP-N},
 we have finally achieved
 our double--Monte Carlo scattering--by--scattering simulation of
 the 3-D elastic WIMP--nucleus scattering process
 and can provide
 the 3-D recoil direction and then the recoil energy of
 the WIMP--scattered target nuclei
 event by event
 in different celestial coordinate systems.
 In this paper,
 we focus at first on
 the simulation results of
 the angular distributions of
 the recoil direction (flux)
 as well as
 the accumulated and the average recoil energies of
 the target nuclei
 scattered by incident halo WIMPs
 (circled in the simulation workflow
  in Fig.~\ref{fig:workflow}).
 An investigation on
 the 3-D {\em effective} velocity and kinetic energy distributions of
 halo WIMPs
 {\em scattering off} target nuclei
 in an underground detector
 (indicated by the upper solid blue arrow
  in Fig.~\ref{fig:workflow})
 will be presented
 in Ref.~\cite{DMDDD-fv_eff}
 separately.

 The remainder of this paper is organized as follows.
 In Sec.~2,
 we describe
 the overall workflow of
 our double--Monte Carlo
 scattering--by--scattering simulation procedure of
 3-dimensional elastic WIMP--nucleus scattering
 as well as
 introduce
 an incoming--WIMP coordinate system
 and discuss
 the validation criterion of
 our MC simulation of
 3-D WIMP scattering events.
 Then,
 in Secs.~3, 4, and 5,
 we demonstrate
 the angular distributions of
 the recoil direction (flux)
 as well as
 the accumulated and the average recoil energies of
 scattered target nuclei
 observed in the incoming--WIMP,
 the laboratory--dependent laboratory,
 and the laboratory--independent Equatorial
 coordinate systems,
 respectively.
 Their target and WIMP--mass dependences
 as well as
 the annual modulations
 will also be discussed
 in detail.
 We conclude in Sec.~6.

%
 \input{sec-toolbox-NR_gen-chi}
%

%
 \input{sec-NR-chi}
%

%
 \input{sec-NR_ang-Lab}
%

%
 \input{sec-NR_ang-Eq}
\section{Summary}

 In this paper,
 as the second part of the third step of
 our development of
 data analysis procedures
 for using and/or combining 3-dimensional information
 offered by directional direct Dark Matter detection experiments
 in the future,
 we investigated
 the angular distributions of
 the recoil direction (flux)
 and the recoil energy of
 the Monte Carlo simulated WIMP--scattered target nuclei
 observed in different celestial coordinate systems.

 At first,
 we demonstrated
 the angular (all--sky)
 and the recoil--angle distributions of
 the recoil direction (flux)
 as well as
 the accumulated and the average recoil energies of
 several frequently used target nuclei
 in the incoming--WIMP coordinate system.
 Our simulations showed that,
 firstly,
 while
 the angular distributions of
 the recoil direction
 and the accumulated/average recoil energies of
 all simulated target nucleus--WIMP mass combinations
 show the azimuthal symmetry
 in the incoming--WIMP point of view,
 their recoil--angle distributions
 have clearly
 strong target and WIMP--mass dependences.
 It would be important to emphasize here that
 the most frequent and the most energetic recoil directions (angles)
 do {\em not} appear
 around the incoming direction of incident WIMPs at all,
 but have the upper bounds of 45$^{\circ}$
 and 60$^{\circ}$,
 respectively,
 and
 reduce strongly
 with the increased nuclear and/or WIMP mass.

 Secondly,
 the obvious discrepancies between
 the simulated recoil--angle distributions of
 the recoil direction/energy
 and the theoretical expectations
 for scattering by WIMPs
 carrying monotonically
 (the half of) the average kinetic energy
 indicate clearly that,
 the kinetic behavior of
 the subgroup of WIMPs
 scattering off (specified) target nuclei
 in an underground detector
 would be different from that of
 the entire incident halo WIMPs:
 due to the sharply reduced
 cross section (nuclear form factor)
 with the increased recoil energy,
 the event numbers
 as well as
 the accumulated and the average recoil energies
 transferred by WIMPs
 moving with high incident velocity
 and/or in large recoil angles
 would be strongly suppressed.

 Meanwhile,
 the angular recoil--direction/energy distributions of
 several target nuclei
 in the laboratory coordinate system
 were also presented
 in this paper,
 since
 this 3-D information,
 combined with the recoil spectrum,
 could practically be measured
 in directional direct DM detection experiments.
 Our simulations showed that,
 on one hand,
 for one given target nucleus,
 the heavier its mass,
 the more obvious the (qualitative) differences
 between the angular distribution patterns
 simulated with different WIMP masses;
 on the other hand,
 the larger the mass difference
 between two target nuclei
 (and the larger the WIMP mass),
 the larger
 the pattern differences
 between their angular distributions.
 This indicates that,
 by comparing and/or combining
 the (quantitative) nuclear recoil flux and energy
 or even only their (qualitative) distribution patterns
 with the same/different target nuclei
 (provided from different underground laboratories),
 one could constrain/identify the mass (range) of halo WIMPs.

 Furthermore,
 since,
 considering the very low theoretically estimated event rate,
 combining and analyzing WIMP scattering events
 provided with the same target nucleus
 from different underground laboratories
 in a laboratory/location--independent coordinate system
 would be practically a very useful strategy,
 we also demonstrated
 the angular distributions of
 the recoil direction/energy
 in the Equatorial coordinate system.
 While
 the anisotropies of
 the distribution patterns of
 the recoil direction/energy
 could be observed clearly,
 neither
 the centers of the most frequent recoil directions
 nor
 those of the most energetic ones of
 all simulated target nucleus--WIMP mass combinations
 could match
 the theoretical main direction of
 the WIMP wind,
 but have clearly northerly shifts,
 which increase
 with the increased mass of the target nucleus
 and/or incident WIMPs.
 Interestingly,
 the hot--points of
 the angular distributions of
 the average recoil energy
 seem to be approximately centered
 around the theoretical main direction of
 incident WIMPs,
 with nevertheless
 some small (but observable)
 target/WIMP--mass dependent pattern differences.

 Since
 the original purpose of
 directional direct Dark Matter detection experiments
 is to observe the directionality
 (the diurnal modulated anisotropy) of
 the recoil direction of
 WIMP--nucleus scattering events,
 we demonstrated equivalently
 the annual modulated anisotropies of
 the distribution patterns of
 the recoil direction (flux)/energy of
 the scattered target nuclei
 in the incoming--WIMP and the Equatorial coordinate systems.
 Our simulations showed that,
 firstly,
 more scattered target nuclei
 would go to smaller (larger) recoil angles
 with larger (smaller) recoil energies
 in the advanced summer (winter).
 Secondly,
 while
 the angular distribution patterns of
 the average recoil energy
 show indeed approximately circular clockwise rotations
 following the instantaneous theoretical main direction of
 the incident halo WIMPs,
 the approximate rotation centers of
 the distribution patterns of
 the recoil direction and
 the accumulated recoil energy
 would clearly shift northwesterly.
 Moreover,
 the annual variations of
 the angular recoil--direction distributions of
 middle--mass and heavy target nuclei
 seem not to be circular,
 but oscillated somehow latitudinally.
 And the shapes of the most frequent recoil directions
 seem also to be somehow longitudinally squeezed;
 the heavier the nuclear (and/or the WIMP) mass,
 the stronger the pattern distortion would be.
 So far
 we have no reasonable explanation(s)
 for this (unexpected) observations.

 In our simulations presented in this paper,
 500 accepted WIMP--scattering events
 on average
 in one observation period
 (365 days/year
  or 60 days/season)
 in one experiment
 for one laboratory/target nucleus
 have been simulated.
 Regarding the observation periods
 considered in our simulations
 presented in this paper,
 we used several approximations
 about the Earth's orbital motion
 in the Solar system.
 First,
 the Earth's orbit around the Sun is perfectly circular
 on the Ecliptic plane
 and the orbital speed is thus a constant.
 Second,
 the date of the vernal equinox is exactly fixed
 at the end of the May 20th (the 79th day) of
 a 365-day year
 and the few extra hours
 in an actual Solar year
 have been neglected.
 Nevertheless,
 considering the very low WIMP scattering event rate
 and thus maximal a few (tens) of total (combined) WIMP events
 observed in at least a few tens (or even hundreds) of days
 (an optimistic overall event rate of $\lsim \~ {\cal O}(1)$ event/day)
 for the first--phase analyses,
 these approximations should be acceptable.

 In summary,
 we finally achieved
 the full Monte Carlo scattering--by--scattering simulation of
 the 3-dimensional elastic WIMP--nucleus scattering process
 and can provide experimentally measurable (pseudo)data:
 the 3-dimensional recoil direction
 and the recoil energy of
 the WIMP--scattered target nuclei.
 Several important (but unexpected) characteristics
 have been observed.
 Hopefully,
 this (and more works fulfilled in the future)
 could help our colleagues
 to develop analysis methods
 for understanding
 the astrophysical and particle properties of
 Galactic WIMPs
 as well as
 the structure of Dark Matter halo
 by using directional direct detection data.

\subsubsection*{Acknowledgments}

 The author would like to thank
 the pleasant atmosphere of
 the W101 Ward and the Cancer Center of
 the Kaohsiung Veterans General Hospital,
 where part of this work was completed.
 This work
 was strongly encouraged by
 the ``{\it Researchers working on
 e.g.~exploring the Universe or landing on the Moon
 should not stay here but go abroad.}'' speech.

%
%
%
 \input{sec-references}
%

%
%
%
\end{document}

%% file: def-newcommands.tex
 \def \PlotNumberAa {00000}
 \def \PeriodCa     {19.49  --  79.49}
 \def \PeriodCb     {110.74 -- 170.74}
 \def \PeriodCc     {201.99 -- 261.99}
 \def \PeriodCd     {293.24 -- 353.24}
 \def \PlotNumberCa {04949}
 \def \PlotNumberCb {14074}
 \def \PlotNumberCc {23199}
 \def \PlotNumberCd {32324}
\newcommand{\rmF}   {\rmXA{F}  {19}}
\newcommand{\rmGe}  {\rmXA{Ge} {73}}
\newcommand{\rmXe}  {\rmXA{Xe}{129}}
\newcommand{\rmW}   {\rmXA{W} {183}}
\newcommand{\InsertSKPPlotS} [3] [10.5] {
\begin{figure} [t!]
\begin{center}
 \includegraphics [width = #1 cm] {skp-#2}
\end{center}
\caption{
 #3
}
\label{fig:#2}
\end{figure}
}
\newcommand{\OnlinePlotNR} [8] {
\href{http://www.tir.tw/phys/hep/dm/amidas-2d/amidas-2d.php%
      ?amidas_2D_function=#1%
      &mode_NR=#3%
      &frame=#4%
      &mode_animation=#5%
      &period=#6%
      &event_No=#7}
     {#8}%
}
\newcommand{\InsertPlotNRchi} [3] {
\begin{figure} [t!]
\begin{center}
 \begin{subfigure} [c] {8.4 cm}%
  \OnlinePlotNR
   {NR_ang}
   {} {NR_ang}
   {\ShortFrame}
   {#1}
   {periodA}
   {#2}
   {\includegraphics [width = 8.4 cm]
     {NR_ang-\Target-\WIMPmass-\ShortFrame-\EventNumber-\PlotNumber}}%
 \end{subfigure}
 \begin{subfigure} [c] {8.4 cm}%
  \OnlinePlotNR
   {NR_theta}
   {} {NR_theta}
   {\ShortFrame}
   {#1}
   {periodA}
   {#2}
   {\includegraphics [width = 8.4 cm]
     {NR_theta-\Target-\WIMPmass-\ShortFrame-\EventNumber-\PlotNumber}}%
 \end{subfigure}
 \\ \vspace{0.6 cm}
 \begin{subfigure} [c] {8.4 cm}%
  \OnlinePlotNR
   {NR_ang}
   {} {Q_ang}
   {\ShortFrame}
   {#1}
   {periodA}
   {#2}
   {\includegraphics [width = 8.4 cm]
     {Q_ang-\Target-\WIMPmass-\ShortFrame-\EventNumber-\PlotNumber}}%
 \end{subfigure}
 \begin{subfigure} [c] {8.4 cm}%
  \OnlinePlotNR
   {NR_theta}
   {} {Q_theta}
   {\ShortFrame}
   {#1}
   {periodA}
   {#2}
   {\includegraphics [width = 8.4 cm]
     {Q_theta-\Target-\WIMPmass-\ShortFrame-\EventNumber-\PlotNumber}}%
 \end{subfigure}
 \\ \vspace{0.6 cm}
 \begin{subfigure} [c] {8.4 cm}%
  \OnlinePlotNR
   {NR_ang}
   {} {QoN_ang}
   {\ShortFrame}
   {#1}
   {periodA}
   {#2}
   {\includegraphics [width = 8.4 cm]
     {QoN_ang-\Target-\WIMPmass-\ShortFrame-\EventNumber-\PlotNumber}}%
 \end{subfigure}
 \begin{subfigure} [c] {8.4 cm}%
  \OnlinePlotNR
   {NR_theta}
   {} {QoN_theta}
   {\ShortFrame}
   {#1}
   {periodA}
   {#2}
   {\includegraphics [width = 8.4 cm]
     {QoN_theta-\Target-\WIMPmass-\ShortFrame-\EventNumber-\PlotNumber}}%
 \end{subfigure}
\end{center}
\caption{
 #3
}
\label{fig:NR-\Target-\WIMPmass-\ShortFrame-\EventNumber-\PlotNumber}
\end{figure}
}
\newcommand{\InsertPlotNRAnnual} [2] {
\begin{figure} [t!]
\begin{center}
 \OnlinePlotNR
  {NR_#1}
  {} {NR_#1}
  {\ShortFrame}
  {annual&target=\Target&mchi=100}
  {periodC}
  {500}
  {\begin{subfigure} [c] {4.2 cm}%
    \includegraphics [width = 4.2 cm]
     {NR_#1-\Target-\WIMPmass-\ShortFrame-\EventNumber-\PlotNumbera}%
   \end{subfigure}
   \begin{subfigure} [c] {4.2 cm}%
    \includegraphics [width = 4.2 cm]
     {NR_#1-\Target-\WIMPmass-\ShortFrame-\EventNumber-\PlotNumberb}%
   \end{subfigure}
   \begin{subfigure} [c] {4.2 cm}%
    \includegraphics [width = 4.2 cm]
     {NR_#1-\Target-\WIMPmass-\ShortFrame-\EventNumber-\PlotNumberc}%
   \end{subfigure}
   \begin{subfigure} [c] {4.2 cm}%
    \includegraphics [width = 4.2 cm]
     {NR_#1-\Target-\WIMPmass-\ShortFrame-\EventNumber-\PlotNumberd}%
   \end{subfigure}}
 \\ \vspace{0.6 cm}
 \OnlinePlotNR
  {NR_#1}
  {} {Q_#1}
  {\ShortFrame}
  {annual&target=\Target&mchi=100}
  {periodC}
  {500}
  {\begin{subfigure} [c] {4.2 cm}%
    \includegraphics [width = 4.2 cm]
     {Q_#1-\Target-\WIMPmass-\ShortFrame-\EventNumber-\PlotNumbera}%
   \end{subfigure}
   \begin{subfigure} [c] {4.2 cm}%
    \includegraphics [width = 4.2 cm]
     {Q_#1-\Target-\WIMPmass-\ShortFrame-\EventNumber-\PlotNumberb}%
   \end{subfigure}
   \begin{subfigure} [c] {4.2 cm}%
    \includegraphics [width = 4.2 cm]
     {Q_#1-\Target-\WIMPmass-\ShortFrame-\EventNumber-\PlotNumberc}%
   \end{subfigure}
   \begin{subfigure} [c] {4.2 cm}%
    \includegraphics [width = 4.2 cm]
     {Q_#1-\Target-\WIMPmass-\ShortFrame-\EventNumber-\PlotNumberd}%
   \end{subfigure}}
 \\ \vspace{0.6 cm}
 \OnlinePlotNR
  {NR_#1}
  {} {QoN_#1}
  {\ShortFrame}
  {annual&target=\Target&mchi=100}
  {periodC}
  {500}
  {\begin{subfigure} [c] {4.2 cm}%
    \includegraphics [width = 4.2 cm]
     {QoN_#1-\Target-\WIMPmass-\ShortFrame-\EventNumber-\PlotNumbera}%
   \end{subfigure}
   \begin{subfigure} [c] {4.2 cm}%
    \includegraphics [width = 4.2 cm]
     {QoN_#1-\Target-\WIMPmass-\ShortFrame-\EventNumber-\PlotNumberb}%
   \end{subfigure}
   \begin{subfigure} [c] {4.2 cm}%
    \includegraphics [width = 4.2 cm]
     {QoN_#1-\Target-\WIMPmass-\ShortFrame-\EventNumber-\PlotNumberc}%
   \end{subfigure}
   \begin{subfigure} [c] {4.2 cm}%
    \includegraphics [width = 4.2 cm]
     {QoN_#1-\Target-\WIMPmass-\ShortFrame-\EventNumber-\PlotNumberd}%
   \end{subfigure}}
 \\ \vspace{0.3 cm}
   \begin{subfigure} [c] {4.2 cm}%
   \caption{\footnotesize \Perioda\ day}
   \end{subfigure}
   \begin{subfigure} [c] {4.2 cm}%
   \caption{\footnotesize \Periodb\ day}
   \end{subfigure}
   \begin{subfigure} [c] {4.2 cm}%
   \caption{\footnotesize \Periodc\ day}
   \end{subfigure}
   \begin{subfigure} [c] {4.2 cm}%
   \caption{\footnotesize \Periodd\ day}
   \end{subfigure}
 \\ \vspace{-0.45 cm}
\end{center}
\caption{
 #2
}
\label{fig:NR_#1-\Target-\WIMPmass-\ShortFrame-\EventNumber-\PlotNumbera}
\end{figure}
}
\newcommand{\InsertPlotNRmchi} [1] {
\begin{figure} [t!]
\begin{center}
 \OnlinePlotNR
  {NR_ang}
  {} {NR_ang}
  {\ShortFrame}
  {mchi&target=\Target}
  {periodA}
  {500}
  {\begin{subfigure} [c] {5.5 cm}%
    \includegraphics [width = 5.5 cm]
     {NR_ang-\Target-0020-\ShortFrame-\EventNumber-\PlotNumber}%
   \end{subfigure}
   \begin{subfigure} [c] {5.5 cm}%
    \includegraphics [width = 5.5 cm]
     {NR_ang-\Target-0100-\ShortFrame-\EventNumber-\PlotNumber}%
   \end{subfigure}
   \begin{subfigure} [c] {5.5 cm}%
    \includegraphics [width = 5.5 cm]
     {NR_ang-\Target-0200-\ShortFrame-\EventNumber-\PlotNumber}%
   \end{subfigure}}
 \\ \vspace{0.6 cm}
 \OnlinePlotNR
  {NR_ang}
  {} {Q_ang}
  {\ShortFrame}
  {mchi&target=\Target}
  {periodA}
  {500}
  {\begin{subfigure} [c] {5.5 cm}%
    \includegraphics [width = 5.5 cm]
     {Q_ang-\Target-0020-\ShortFrame-\EventNumber-\PlotNumber}%
   \end{subfigure}
   \begin{subfigure} [c] {5.5 cm}%
    \includegraphics [width = 5.5 cm]
     {Q_ang-\Target-0100-\ShortFrame-\EventNumber-\PlotNumber}%
   \end{subfigure}
   \begin{subfigure} [c] {5.5 cm}%
    \includegraphics [width = 5.5 cm]
     {Q_ang-\Target-0200-\ShortFrame-\EventNumber-\PlotNumber}%
   \end{subfigure}}
 \\ \vspace{0.6 cm}
 \OnlinePlotNR
  {NR_ang}
  {} {QoN_ang}
  {\ShortFrame}
  {mchi&target=\Target}
  {periodA}
  {500}
  {\begin{subfigure} [c] {5.5 cm}%
    \includegraphics [width = 5.5 cm]
     {QoN_ang-\Target-0020-\ShortFrame-\EventNumber-\PlotNumber}%
   \end{subfigure}
   \begin{subfigure} [c] {5.5 cm}%
    \includegraphics [width = 5.5 cm]
     {QoN_ang-\Target-0100-\ShortFrame-\EventNumber-\PlotNumber}%
   \end{subfigure}
   \begin{subfigure} [c] {5.5 cm}%
    \includegraphics [width = 5.5 cm]
     {QoN_ang-\Target-0200-\ShortFrame-\EventNumber-\PlotNumber}%
   \end{subfigure}}
 \\ \vspace{0.3 cm}
   \begin{subfigure} [c] {5.5 cm}%
   \caption{$\mchi =  20$ GeV}
   \end{subfigure}
   \begin{subfigure} [c] {5.5 cm}%
   \caption{$\mchi = 100$ GeV}
   \end{subfigure}
   \begin{subfigure} [c] {5.5 cm}%
   \caption{$\mchi = 200$ GeV}
   \end{subfigure}
 \\ \vspace{-0.45 cm}
\end{center}
\caption{
 #1
}
\label{fig:NR-\Target-\ShortFrame-\EventNumber-\PlotNumber}
\end{figure}
}
\newcommand{\InsertPlotNRLabmchi} [1] {
\begin{figure} [t!]
\begin{center}
 \OnlinePlotNR
  {NR_ang}
  {} {NR_ang}
  {\ShortFrame&ULab=\LabName}
  {mchi&target=\Target}
  {periodA}
  {500}
  {\begin{subfigure} [c] {5.5 cm}%
    \includegraphics [width = 5.5 cm]
     {NR_ang-\Target-0020-\ShortFrame-\EventNumber-\PlotNumber-\LabName}%
   \end{subfigure}
   \begin{subfigure} [c] {5.5 cm}%
    \includegraphics [width = 5.5 cm]
     {NR_ang-\Target-0100-\ShortFrame-\EventNumber-\PlotNumber-\LabName}%
   \end{subfigure}
   \begin{subfigure} [c] {5.5 cm}%
    \includegraphics [width = 5.5 cm]
     {NR_ang-\Target-0200-\ShortFrame-\EventNumber-\PlotNumber-\LabName}%
   \end{subfigure}}
 \\ \vspace{0.6 cm}
 \OnlinePlotNR
  {NR_ang}
  {} {Q_ang}
  {\ShortFrame&ULab=\LabName}
  {mchi&target=\Target}
  {periodA}
  {500}
  {\begin{subfigure} [c] {5.5 cm}%
    \includegraphics [width = 5.5 cm]
     {Q_ang-\Target-0020-\ShortFrame-\EventNumber-\PlotNumber-\LabName}%
   \end{subfigure}
   \begin{subfigure} [c] {5.5 cm}%
    \includegraphics [width = 5.5 cm]
     {Q_ang-\Target-0100-\ShortFrame-\EventNumber-\PlotNumber-\LabName}%
   \end{subfigure}
   \begin{subfigure} [c] {5.5 cm}%
    \includegraphics [width = 5.5 cm]
     {Q_ang-\Target-0200-\ShortFrame-\EventNumber-\PlotNumber-\LabName}%
   \end{subfigure}}
 \\ \vspace{0.6 cm}
 \OnlinePlotNR
  {NR_ang}
  {} {QoN_ang}
  {\ShortFrame&ULab=\LabName}
  {mchi&target=\Target}
  {periodA}
  {500}
  {\begin{subfigure} [c] {5.5 cm}%
    \includegraphics [width = 5.5 cm]
     {QoN_ang-\Target-0020-\ShortFrame-\EventNumber-\PlotNumber-\LabName}%
   \end{subfigure}
   \begin{subfigure} [c] {5.5 cm}%
    \includegraphics [width = 5.5 cm]
     {QoN_ang-\Target-0100-\ShortFrame-\EventNumber-\PlotNumber-\LabName}%
   \end{subfigure}
   \begin{subfigure} [c] {5.5 cm}%
    \includegraphics [width = 5.5 cm]
     {QoN_ang-\Target-0200-\ShortFrame-\EventNumber-\PlotNumber-\LabName}%
   \end{subfigure}}
 \\ \vspace{0.3 cm}
   \begin{subfigure} [c] {5.5 cm}%
   \caption{$\mchi =  20$ GeV}
   \end{subfigure}
   \begin{subfigure} [c] {5.5 cm}%
   \caption{$\mchi = 100$ GeV}
   \end{subfigure}
   \begin{subfigure} [c] {5.5 cm}%
   \caption{$\mchi = 200$ GeV}
   \end{subfigure}
 \\ \vspace{-0.45 cm}
\end{center}
\caption{
 #1
}
\label{fig:NR-\Target-\ShortFrame-\EventNumber-\PlotNumber-\LabName}
\end{figure}
}

%% file: sec-toolbox-NR_gen-chi.tex
%
%
\section{Monte Carlo scattering--by--scattering simulation of
         3-dimensional elastic WIMP--nucleus scattering events}
\label{sec:3D-WIMP-N}

 In Ref.~\cite{DMDDD-3D-WIMP-N},
 we have
\begin{enumerate}
\item
 reviewed
 the MC generation of
 the 3-D velocity information
 (the magnitude,
  the direction,
  and
  the incoming/scattering time)
 of Galactic WIMPs,
\item
 summarized
 the definitions of
 and
 the transformations between
 all celestial coordinate systems.
\end{enumerate}
 In this section,
 we focus then on the core part of
 our simulation procedure:
 the generation and the validation of
 3-D elastic WIMP--nucleus scattering events
 in the ``incoming--WIMP'' coordinate system.

 We describe at first
 the overall workflow of
 our Monte Carlo scattering--by--scattering simulation of
 the 3-D elastic WIMP--nucleus scattering process.
 Then
 we give
 our definition of
 the incoming--WIMP coordinate system
 as well as
 those of
 (the orientation of) the scattering plane
 and the (equivalent) recoil angle.
 At the end
 we discuss
 the validation criterion of
 our Monte Carlo simulations
 by taking into account
 the cross section (nuclear form factor) suppression
 on {\em each} generated recoil energy.

\subsection{Simulation workflow}
\label{sec:workflow}
\begin{figure} [t!]
\begin{center}
 \includegraphics [width = 15.5 cm] {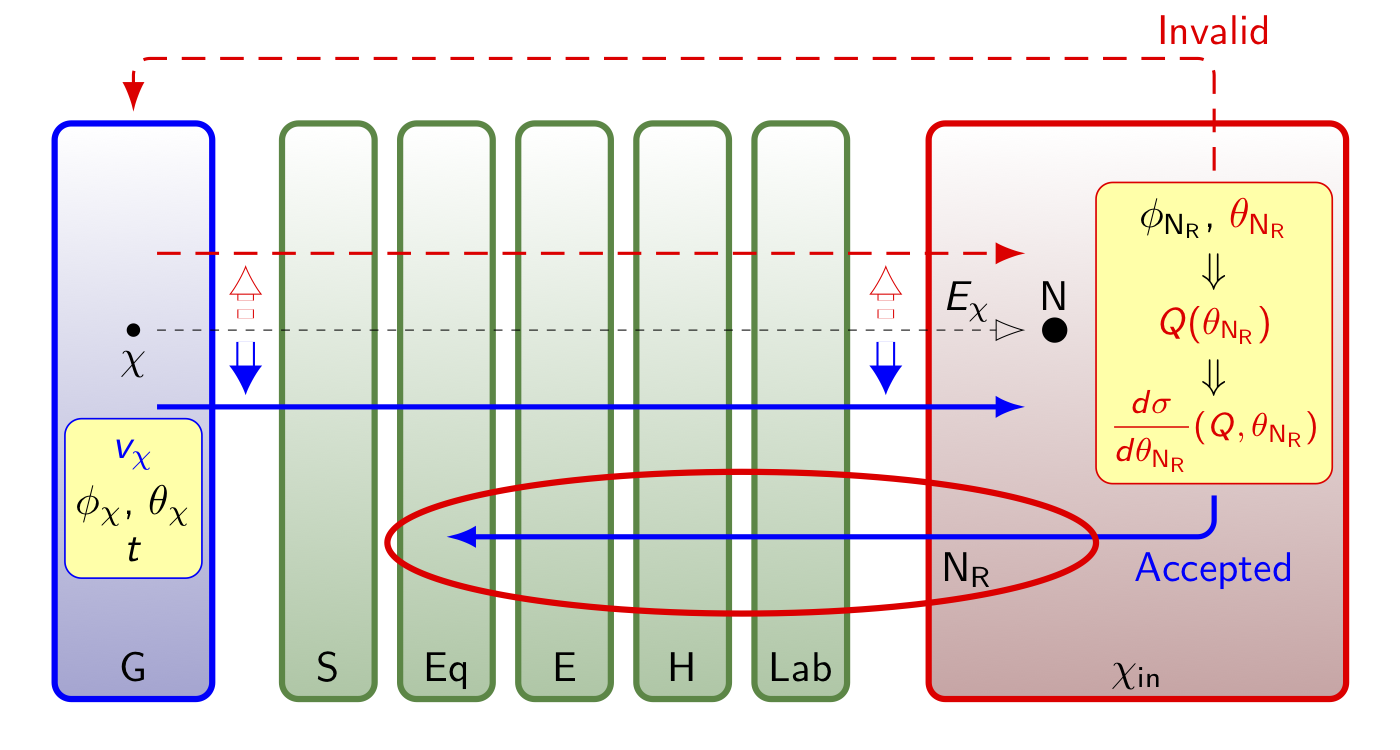}
\end{center}
\caption{
 The workflow of
 our double--Monte Carlo simulation and data analysis procedure of
 3-dimensional elastic WIMP--nucleus scattering.
 See the text for detailed descriptions.
}
\label{fig:workflow}
\end{figure}

 In this subsection,
 we describe
 the overall workflow of
 our double--Monte Carlo simulation and data analysis procedure of
 3-D elastic WIMP--nucleus scattering
 sketched in Fig.~\ref{fig:workflow} in detail:
\begin{enumerate}
\item
 The 3-D velocity information of incident halo WIMPs
 (the magnitude and the direction
  as well as
  the incoming/scattering time)
 is MC generated
 according to a specified model of the Dark Matter halo
 in the Galactic coordinate system
 (the blue subframe)
 \cite{DMDDD-N, DMDDD-3D-WIMP-N}.
\item
 The generated 3-D WIMP velocities
 will be transformed through
 the laboratory--independent
 (Ecliptic,
  Equatorial,
  and Earth)
 coordinate systems
 as well as
 the laboratory--dependent
 (horizontal and laboratory)
 coordinate systems
 (the green subframes)
 \cite{DMDDD-N, DMDDD-3D-WIMP-N}
 and at the end into the incoming--WIMP coordinate system
 (the red subframe,
  defined in Sec.~\ref{sec:XYZ_chi}).
\item
 In the incoming--WIMP coordinate system,
 the 3-D elastic WIMP--nucleus scattering process
 will also be MC simulated
 by generating
 the orientation of the scattering plane $\phiNRchi$
 and the ``equivalent'' recoil angle $\thetaNRchi$
 (defined in Sec.~\ref{sec:phi_theta_NR_chi}).
 They define the recoil direction of
 the scattered target nucleus
 and the latter,
 combined with the transformed WIMP incident velocity,
 will then be used for estimating
 the transferred recoil energy to
 the target nucleus,
 $Q(\thetaNRchi)$,
 and the differential WIMP--nucleus scattering cross section
 with respect to
 the recoil angle,
 $d\sigma / d\thetaNRchi (Q, \thetaNRchi)$,
 in our event validation criterion
 (see Sec.~\ref{sec:phi_theta_NR_chi}
  for details).
\item
 The orientation of
 the scattering plane $\phiNRchi$
 and the equivalent recoil angle $\thetaNRchi$
 of the {\em accepted} recoil events
 will be transformed (back)
 through all considered celestial coordinate systems
 (indicated by the lower solid blue arrow).
 All these 3-D recoil information of
 the scattered target nucleus
 accompanied with
 the corresponding recoil energy $Q$
 as well as
 the 3-D velocity of the scattering WIMP
 in different coordinate systems
 (the upper solid blue arrow)
 will be recorded
 for further analyses%
\footnote{
 In this paper,
 we focus on investigating
 the angular distributions of
 the nuclear recoil direction (flux)/energy
 (indicated by the lower blue arrow).
 A detailed study on
 the 3-dimensional effective velocity distribution of
 the incident WIMPs
 scattering off target nuclei
 (the upper blue arrow)
 is presented
 in Ref.~\cite{DMDDD-fv_eff}
 separately.
}.
\item
 For the {\em invalid} cases,
 in which
 the estimated recoil energies
 are out of the experimental measurable energy window
 or suppressed by the validation criterion,
 the generated 3-D information
 on the incident WIMP
 (the lower dashed red arrow)
 (and that on the scattered nucleus)
 will be discarded
 and the generation/validation process of
 one WIMP scattering event
 will be restarted from the Galactic coordinate system
 (the upper dashed red arrow).
\end{enumerate}
\InsertSKPPlotS
 {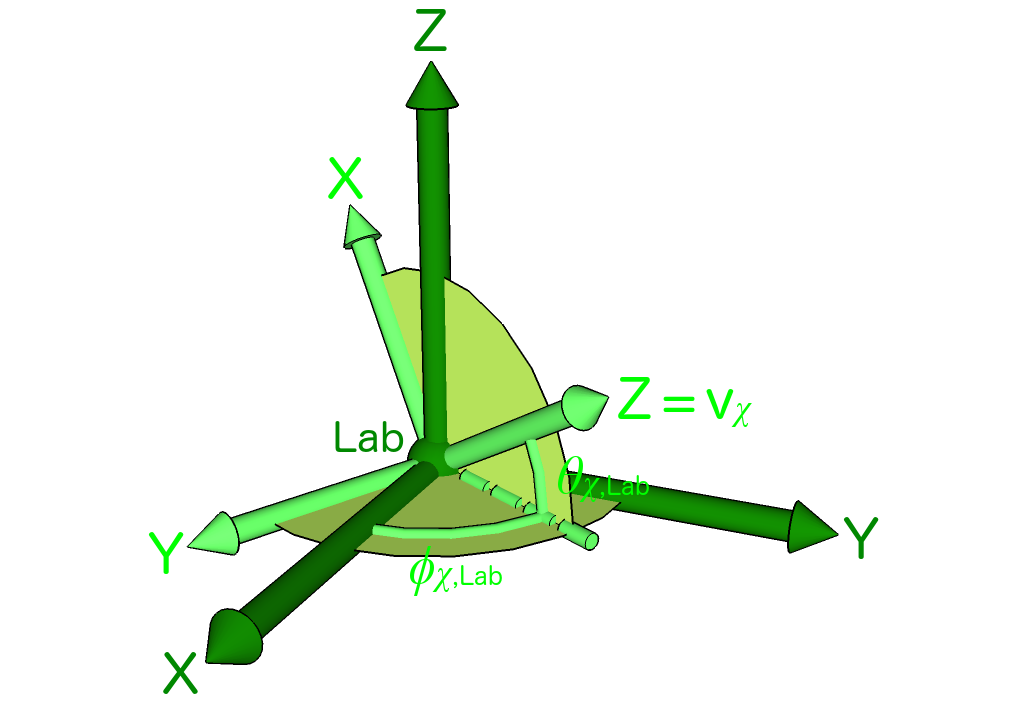}
 {The definition of
  the (light--green) incoming--WIMP coordinate system
  in the (dark--green) laboratory coordinate system.
  The $\zchi$--axis is defined as usual as
  the direction of the incident velocity of
  the incoming WIMP
  $\Vchi$.
  $\phichiLab$ and $\thetachiLab$ indicate
  the azimuthal angle and the elevation of
  the direction of
  $\Vchi$
  measured in the laboratory coordinate system,
  respectively.
  The $\xchi$--axis is perpendicular to the $\zchi$--axis
  and lies on the $\zLab$--$\zchi$ plane.
  Then
  the $\ychi$--axis is defined
  by the right--handed convention.%
  }
\subsection{Definition of the incoming--WIMP coordinate system}
\label{sec:XYZ_chi}

 In Fig.~\ref{fig:chi-Lab},
 we sketch the definition of
 the (light--green) incoming--WIMP coordinate system
 in the (dark--green) laboratory coordinate system.
 Note that,
 practically,
 the center of the incoming--WIMP coordinate system
 is at the position of the scattered target nucleus
 before scattering
 (see Fig.~\ref{fig:NR-chi-Lab}).
 The $\zchi$--axis is defined as usual as
 the direction of the incident velocity of
 the incoming WIMP
 $\Vchi$.
 $\phichiLab$ and $\thetachiLab$ indicate
 the azimuthal angle and the elevation of
 the direction of
 $\Vchi$
 measured in the laboratory coordinate system,
 respectively.
 The $\xchi$--axis is perpendicular to the $\zchi$--axis
 and lies on the $\zLab$--$\zchi$ plane.
 Then
 the $\ychi$--axis is defined
 by the right--handed convention.

 Note that,
 in our Monte Carlo simulations of
 3-D elastic WIMP--nucleus scattering,
 the velocity (incident direction) of halo WIMPs
 in the laboratory
 and
 the Equatorial coordinate systems
 as well as
 the $\zchi$--axis of
 the incoming--WIMP coordinate system
 are {\em not fixed}
 as from the direction of the CYGNUS constellation.
 Interested readers can refer to Ref.~\cite{DMDDD-N}
 for the detailed discussions about
 (the annual and the diurnal modulations of)
 the anisotropy of
 the angular distributions of
 the 3-D WIMP velocity (flux)
 in the laboratory
 and the Equatorial coordinate systems.

\subsection{Generation of nuclear recoil directions}
\label{sec:phi_theta_NR_chi}
\InsertSKPPlotS
 {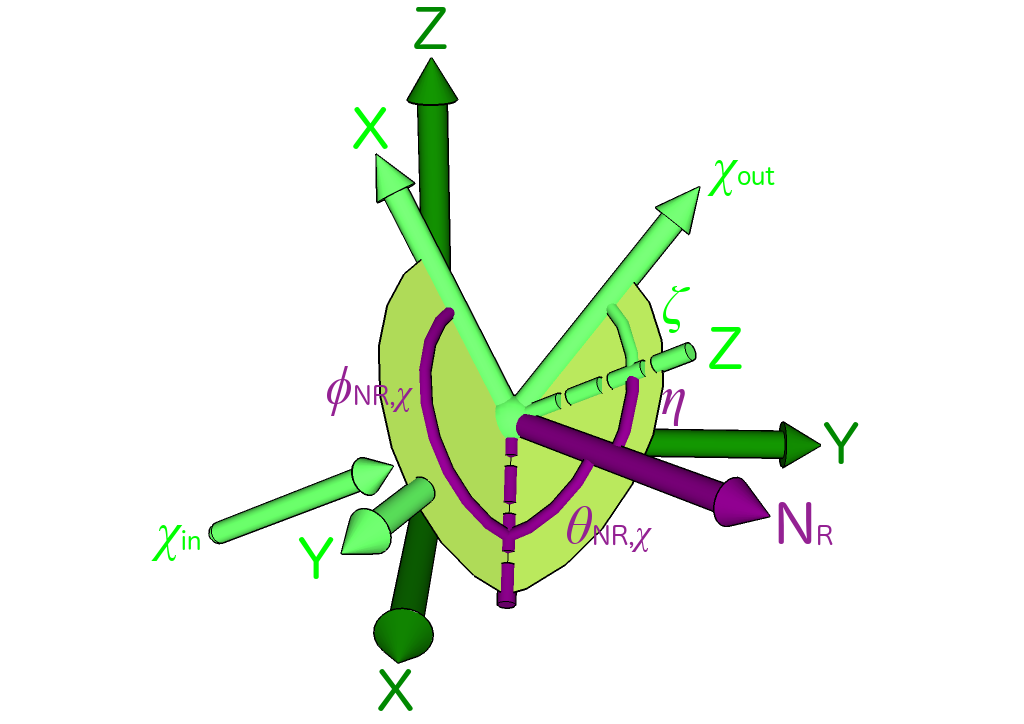}
 {A 3-D elastic WIMP--nucleus scattering event
  in the (light--green) incoming--WIMP
  and the (dark--green) laboratory coordinate systems.
  $\zeta$ and $\eta$ are
  the scattering angle of
  the outgoing WIMP $\chi_{\rm out}$
  and
  the recoil angle of
  the scattered target nucleus N$_{\rm R}$
  measured in the incoming--WIMP coordinate system
  of {\em this single} scattering event,
  respectively.
  While
  the azimuthal angle of
  the recoil direction of
  the scattered nucleus N$_{\rm R}$
  in this incoming--WIMP coordinate system,
  $\phiNRchi$,
  indicates the orientation of the scattering plane,
  the elevation of
  the recoil direction of N$_{\rm R}$,
  $\thetaNRchi$,
  is namely the complementary angle of
  the recoil angle
  $\eta$.%
  }

 In Fig.~\ref{fig:NR-chi-Lab},
 we sketch 
 the process of
 one single 3-D elastic WIMP--nucleus scattering event:
 $\chi_{\rm in/out}$ indicate
 the incoming and the outgoing WIMPs,
 respectively.
 While
 $\zeta$ indicates
 the scattering angle of
 the outgoing WIMP $\chi_{\rm out}$
 (measured from the $\zchi$--axis),
 $\eta$ is the recoil angle of
 the scattered target nucleus N$_{\rm R}$.

 It can be found firstly that
 the orientation of the ($\Vchiout$--$\zchi$--${\bf v}_{\rm N_R}$) scattering plane
 of {\em this single} scattering event
 can be specified by
 the azimuthal angle of
 the recoil direction of
 the scattered nucleus,
 $\phiNRchi$,
 which
 should be azimuthal symmetric
 around the $\zchi$--axis
 and is thus
 generated with a constant probability
 in our simulations:
\beq
     f_{{\rm N_R}, \chiin, \phi}(\phiNRchi)
  =  1
\~,
     ~~~~ ~~~~ ~~ 
     \phiNRchi \in (-\pi,~\pi]
\~.
\label{eqn:f_NR_phiNRchi}
\eeq

 On the other hand,
 the elevation of
 the recoil direction of
 the scattered nucleus,
 $\thetaNRchi$,
 is namely the complementary angle of
 the recoil angle $\eta$
 and
 we thus use
\beq
      \thetaNRchi
 \in  [0,~\pi / 2]
\label{eqn:thetaNRchi_range}
\eeq
 as the ``equivalent'' recoil angle
 in our simulations%
\footnote{
 Note that,
 without special remark,
 we will use hereafter simply ``the recoil angle''
 to indicate ``the equivalent recoil angle $\thetaNRchi$''
 (not $\eta$).
}.
 Since
 for one WIMP event
 transformed
 into the laboratory coordinate system
 with the velocity of
 $\Vchi(\vchiLab, \phichiLab, \thetachiLab)$,
 the kinetic energy
 can be given by
\beq
     \Echi
  =  \frac{1}{2} \mchi |\Vchi|^2
  =  \frac{1}{2} \mchi \vchiLab^2
\~.
\label{eqn:Echi}
\eeq
 Then
 the recoil energy of the scattered target nucleus
 in the incoming--WIMP coordinate system
 can be estimated by
 the equivalent recoil angle
 $\thetaNRchi$
 as
 \cite{DMDDD-3D-WIMP-N}
\beq
     Q(\thetaNRchi)
  =  \bbrac{\afrac{2 \mrN^2}{\mN} \vchiLab^2}
     \sin^2(\thetaNRchi)
\~,
\label{eqn:QQ_thetaNRchi}
\eeq
 where
\(
         \mrN
 \equiv  \mchi \mN / (\mchi + \mN)
\)
 is the reduced mass of
 the WIMP mass $\mchi$ and
 that of the target nucleus $\mN$.
 And
 the differential cross section $d\sigma$
 given by
 the absolute value of the momentum transfer
 from the incident WIMP to the recoiling target nucleus,
 $q = |{\bf q}| = \sqrt{2 \mN Q}$,
 can be obtained as
 \cite{SUSYDM96, DMDDD-3D-WIMP-N}
\beq
     d\sigma
  =  \frac{1}{\vchiLab^2}
     \afrac{\sigma_0}{4 \mrN^2} F^2(q) \~ dq^2
  =  \sigma_0 F^2(Q(\thetaNRchi))
     \sin(2 \thetaNRchi) \~ d\thetaNRchi
\~.
\label{eqn:dsigma_thetaNRchi}
\eeq
 Hence,
 the differential WIMP--nucleus scattering cross section
 with respect to
 the recoil angle
 $\thetaNRchi$
 can generally be given by
 \cite{DMDDD-3D-WIMP-N}%
\footnote{
 It would be important to emphasize here that,
 to the best of our knowledge,
 this should be the first time in literature that
 some constraints on
 the nuclear recoil angle/direction
 caused by
 (elastic) WIMP--nucleus scattering cross sections (nuclear form factors)
 have been considered
 in (3-D) WIMP scattering simulations.
}
\beq
     \Dd{\sigma}{\thetaNRchi}
  =  \bbigg{  \sigmaSI F_{\rm SI}^2(Q(\thetaNRchi))
            + \sigmaSD F_{\rm SD}^2(Q(\thetaNRchi)) }
     \sin(2 \thetaNRchi)
\~.
\label{eqn:dsigma_dthetaNRchi}
\eeq
 Here
 $\sigma_0^{\rm (SI, SD)}$ are
 the spin--independent (SI)/spin--dependent (SD) total cross sections
 ignoring the form factor suppression
 and
 $F_{\rm (SI, SD)}(Q)$ indicate the elastic nuclear form factors
 corresponding to the SI/SD WIMP interactions,
 respectively.

 Finally,
 taking into account
 the proportionality of the WIMP flux
 to the incident velocity,
 the generating probability distribution of
 the recoil angle
 $\thetaNRchi$,
 which is proportional to
 the scattering event rate of
 incident halo WIMPs
 with an incoming velocity $\vchiLab$
 off target nuclei
 going into recoil angles of
 $\thetaNRchi \pm d\thetaNRchi / 2$
 with recoil energies of $Q \pm dQ / 2$,
 can generally be given by
 \cite{DMDDD-3D-WIMP-N}
\beq
     f_{{\rm N_R}, \chiin, \theta}(\thetaNRchi)
  =  \afrac{\vchiLab}{v_{\chi, {\rm cutoff}}}
     \bbigg{  \sigmaSI F_{\rm SI}^2(Q(\thetaNRchi))
            + \sigmaSD F_{\rm SD}^2(Q(\thetaNRchi)) }
     \sin(2 \thetaNRchi)
\~,
\label{eqn:f_NR_thetaNRchi}
\eeq
 where
 $v_{\chi, {\rm cutoff}} \simeq 800$ km/s is
 a cut--off velocity of incident halo WIMPs
 in the laboratory coordinate system.

%% file: sec-NR-chi.tex
%
%
\section{Angular distributions of
         the recoil direction/energy
         in the incoming--WIMP frame}
\label{sec:NR-chi}

 In Refs.~\cite{DMDDD-N} and \cite{DMDDD-P},
 we demonstrated
 the angular distributions of
 the 3-D velocity (flux)
 and the kinetic energy of
 Galactic halo WIMPs
 impinging into (directional) direct DM detectors
 in different celestial coordinate systems,
 which show clearly
 the anisotropy and the directionality
 (the annual and/or the diurnal modulations).
 Since
 what we can practically observe (reconstruct)
 in directional Dark Matter detection experiments
 is the recoil energies
 and the recoil tracks (and the head--tail senses)
 of target nuclei,
 in this and the next two sections,
 we discuss
 the angular distributions of
 the recoil direction (flux)
 as well as
 the accumulated and the average recoil energies of
 target nuclei
 scattered by (simulated) incident Galactic WIMPs
 observed in the incoming--WIMP,
 the laboratory--dependent laboratory,
 and the laboratory--independent Equatorial
 coordinate systems,
 respectively.

 Three spin--sensitive nuclei
 used frequently
 in (directional) direct detection experiments:
 $\rmF$,
 $\rmGe$,
 and $\rmXe$
 have been considered
 as our targets%
\footnote{
 Although Ge and Xe are (so far) not used
 in directional detection experiments,
 for readers' cross reference to Ref.~\cite{DMDDD-fv_eff},
 we present simulations with them here,
 which would show similar results
 as with the $\rmXA{Br}{79/81}$ and $\rmXA{I}{127}$ nuclei,
 respectively.
}.
 Then,
 while
 the SI (scalar) WIMP--nucleon cross section
 has been fixed as $\sigmapSI = 10^{-9}$ pb
 in our simulations
 presented in this paper,
 the effective SD (axial--vector) WIMP--proton/neutron couplings
 have been tuned as $\armp = 0.01$
 and $\armn = 0.7 \armp = 0.007$,
 respectively
 \cite{DMDDD-3D-WIMP-N}.
 So that
 the contributions of
 the SI and the SD WIMP--nucleus cross sections
 (including the corresponding nuclear form factors)
 to the validation criterion (\ref{eqn:f_NR_thetaNRchi})
 are approximately comparable
 for the considered $\rmGe$ and $\rmXe$ target nuclei%
\footnote{
 The mass of the $\rmF$ nucleus is too light.
 Thus,
 with the same simulation setup,
 the SD WIMP--nucleus cross section dominates.
}.

 Moreover,
 in this paper
 we assume simply that
 the experimental threshold energies
 for all considered target nuclei
 are negligible,
 whereas
 the maximal experimental cut--off energy
 has been set as $\Qmax = 100$ keV.
 5,000 experiments with 500 accepted
 events on average
 (Poisson--distributed)
 in one observation period
 (365 days/year
  or 60 days/season)
 in one experiment
 for one laboratory/target nucleus
 have been simulated.

 For readers' reference,
 all simulation plots presented in this paper (and more)
 can be found ``in animation''
 on our online (interactive) demonstration webpage
 \cite{AMIDAS-2D-web}.

\subsection{Angular distributions of
            the recoil direction/energy of light target nuclei}
\label{sec:NR-chi-F}

 In this subsection,
 we present at first
 the angular distributions of
 the recoil direction (flux)
 as well as
 the accumulated and the average recoil energies of
 light target nuclei
 in the incoming--WIMP coordinate system.

 \def \Target      {F19}
 \def \WIMPmass    {0100}
 \def \ShortFrame  {chi}
 \def \EventNumber {0500}
 \def \PlotNumber  {\PlotNumberAa}
 \InsertPlotNRchi
  {target&mchi=100}
  {500}
  {Left:
   the angular distributions of
   the recoil direction (flux) (top),
   the accumulated (middle)
   and the average (bottom) recoil energies of
   $\rmF$
   in unit of the all--sky average values.
   (Note that
    the scale of the color bar
    used in the bottom frame is different
    from that in the top and the middle frames.)
   Right:
   the recoil--angle $\thetaNRchi$ dependences of
   the corresponding event number (top),
   the accumulated (middle)
   and the average (bottom) recoil energies of
   $\rmF$;
   the thin vertical dashed black lines
   indicate the 1$\sigma$ statistical uncertainties.
   The mass of incident WIMPs
   has been set as $\mchi = 100$ GeV.
   500 accepted
   WIMP scattering events on average
   in one entire year
   have been simulated
   and binned into 12 $\times$ 12 bins
   for the azimuthal angle and the elevation (left)
   and 12 bins in the range of $\thetaNRchi = 0$ and 90$^{\circ}$ (right),
   respectively.
   See the text for further details.
   \vspace{-0.9  cm}%
   }

 In the left column
 of Figs.~\ref{fig:NR-F19-0100-chi-0500-\PlotNumberAa},
 we show
 the angular distributions of
 the recoil direction (flux) (top),
 the accumulated (middle)
 and the average (bottom) recoil energies of
 $\rmF$
 in unit of the all--sky average values.
 (Note that
  the scale of the color bar
  used in the bottom frame is different
  from that in the top and the middle frames.)
 Meanwhile,
 the corresponding event number (top),
 the accumulated (middle)
 and the average (bottom) recoil energies of
 $\rmF$
 as functions of the recoil angle $\thetaNRchi$
 have also been provided
 in the right column,
 where
 the thin vertical dashed black lines
 indicate the 1$\sigma$ statistical uncertainties.
 The mass of incident WIMPs
 has been set as $\mchi = 100$ GeV.
 500 accepted
 WIMP scattering events on average
 in one entire year
 have been simulated
 and binned into 12 $\times$ 12 bins
 for the azimuthal angle and the elevation (left)
 and 12 bins in the range of $\thetaNRchi = 0$ and 90$^{\circ}$ (right),
 respectively.

 As a reference,
 in the plots
 shown in the right column,
 we draw also
 the dash--dotted red and
 the dash--double--dotted blue curves
 to indicate two theoretical cases:
 all WIMPs move monotonically
 with the root--mean--square velocity
 \cite{DMDDD-3D-WIMP-N}
\beq
         v_{\rm rms, Lab}^2
  =      \afrac{3}{2} v_0^2 + \ve^2
 \simeq  (355~{\rm km/s})^2
\~,
\label{eqn:v2_rms_sh}
\eeq
 namely,
 carry
 the average kinetic energy
\beq
     \Expv{\Echi}
  =  \afrac{1}{2} \mchi v_{\rm rms, Lab}^2
\label{eqn:expv_Echi}
\~,
\eeq
 and
 all WIMPs
 carry only $\Expv{\Echi} / 2$
 (with the common velocity of $v_{\rm rms, Lab} / \sqrt{2} \simeq$ 250 km/s),
 respectively.
 Additionally,
 the quadruple--dotted green curve
 in the top--right frame
 is the exact $\sin(2 \thetaNRchi)$ curve,
 which indicates the case of
 {\em zero} cross section (nuclear form factor) suppression
 considered in the validation criterion
 (\ref{eqn:f_NR_thetaNRchi}).

 First of all,
 it can be seen that,
 since the $\zchi$--axis
 (of each accepted WIMP--nucleus scattering event)
 is defined as
 the incident direction of the incoming WIMP,
 the angular distributions of
 the recoil direction
 as well as
 the accumulated/average recoil energies (of
 the $\rmF$ target)
 in the left column
 of Figs.~\ref{fig:NR-F19-0100-chi-0500-\PlotNumberAa}
 show not surprisingly the ``azimuthal symmetry''.
 However,
 as one can also find in the right column,
 the most frequent and the most energetic recoil directions (angles)
 do {\em not} appear around the $+\zchi$--axis
 ($\thetaNRchi = 90^{\circ}$),
 but around the angle of
 $\thetaNRchi \simeq 45^{\circ}$
 and $\thetaNRchi \simeq 60^{\circ}$,
 respectively!
 More precisely,
 due to the factor of $\sin(2 \thetaNRchi)$
 in the expression (\ref{eqn:f_NR_thetaNRchi})
 for $f_{{\rm N_R}, \chiin, \theta}(\thetaNRchi)$
 and the proportionality of the recoil energy $Q$
 to $\sin^2(\thetaNRchi)$,
 $\theta_{\rm N_R, \chi_{in}, max} = 45^{\circ}$
 and $\theta_{\rm N_R, \chi_{in}, Qmax} = 60^{\circ}$
 are the {\em upper} bounds of
 the most frequent and the most energetic recoil angles,
 respectively.
 And,
 as we will demonstrate later,
 the most frequent/energetic recoil angles
 move towards (much) smaller $\thetaNRchi$,
 when the mass of the target nucleus and/or incident WIMPs
 becomes heavier.

\begin{figure} [t!]
\begin{center}
 \includegraphics [width = 12 cm] {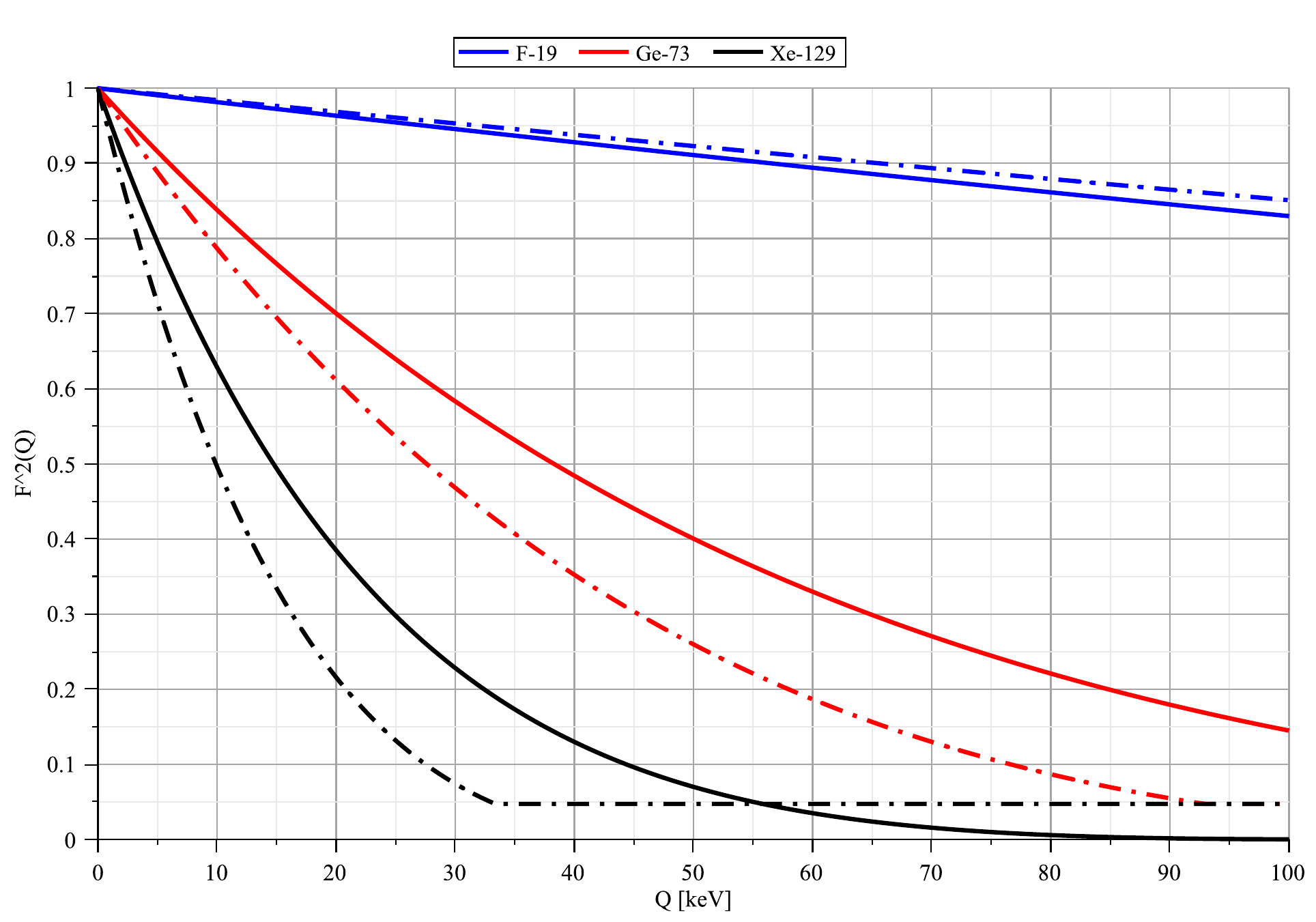}
\end{center}
\caption{
 Nuclear form factors of
 the $\rmF$      (top    blue),
 the $\rmGe$     (middle red),
 and the $\rmXe$ (bottom black) nuclei
 as functions of the recoil energy.
 The solid and dash--dotted curves
 indicate the form factors corresponding to
 the SI and SD cross sections
 adopted in our simulation package,
 respectively
 \cite{DMDDD-3D-WIMP-N}.
}
\label{fig:FQ}
\end{figure}

 Moreover,
 the recoil--angle distribution
 $f_{{\rm N_R}, \chiin, \theta}(\thetaNRchi)$
 of $\rmF$ nuclei
 scattered by 100-GeV WIMPs
 shown in the top--right frame
 of Figs.~\ref{fig:NR-F19-0100-chi-0500-\PlotNumberAa}
 matches almost (but not) perfectly
 the dash--dotted red average--kinetic--energy curve
 (as well as
  the dash--double--dotted blue half--average--kinetic--energy
  and
  the quadruple--dotted green
  zero--form--factor--suppression curves%
\footnote{
 As shown in Fig.~\ref{fig:FQ}
 and will be discussed in detail later,
 this is due to
 the (relatively) pretty flat decrease of
 the scattering cross section (nuclear form factor) of
 the $\rmF$ target
 with the increased recoil energy.
}).
 This seems to indicate that,
 the kinetic behavior of
 (light target nuclei like) $\rmF$
 scattered by (100-GeV) halo WIMPs
 could be approximated by
 scattering by (100-GeV) WIMPs
 moving with monotonically
 the root--mean--square velocity.

 However,
 in the middle-- and bottom--right frames
 of Figs.~\ref{fig:NR-F19-0100-chi-0500-\PlotNumberAa},
 the recoil--angle distributions of
 the accumulated and the average recoil energies of
 the $\rmF$ target
 (scattered by 100-GeV WIMPs),
 namely
 $Q(\thetaNRchi) \~ f_{{\rm N_R}, \chiin, \theta}(\thetaNRchi)$
 and $Q(\thetaNRchi)$,
 respectively,
 indicate that,
 although
 the recoil--angle distributions of
 the recoil energy of
 (light nuclei like) $\rmF$
 scattered by (100-GeV) WIMPs
 could be approximated by
 the (dash--dotted red) average--kinetic--energy curve
 (within the 1$\sigma$ statistical uncertainties),
 discrepancies
 between the theoretical expectations
 and the (black) simulation histograms
 could already be observed
 around the most--energetic
 and in the large--recoil--angle ranges,
 respectively%
\footnote{
 Most (black) simulation histograms here
 are higher than the theoretical expectations.
 In Ref.~\cite{DMDDD-fv_eff},
 we will show that
 the average velocity of
 the (100-GeV) WIMPs
 scattering off (light nuclei like) $\rmF$
 is indeed a bit higher than
 that of
 the entire incident halo WIMPs.
}.
\subsection{Target dependence of
            the angular recoil--direction/energy distributions}
\label{sec:NR-chi-target}
\begin{figure} [t!]
\begin{center}
 \includegraphics [width = 12 cm] {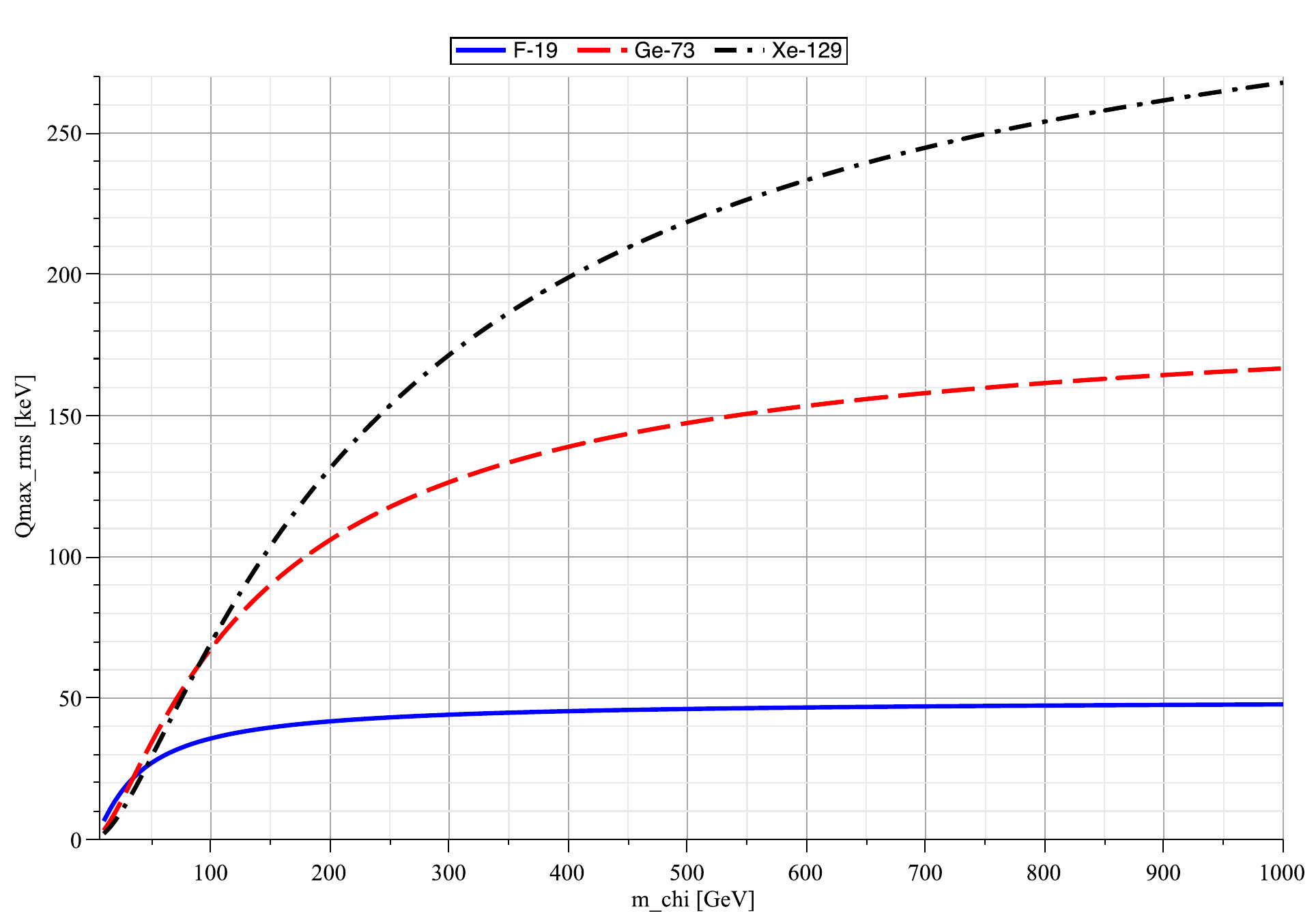}
\end{center}
\caption{
 The WIMP--mass dependence of
 the maximum of the recoil energy,
 $Q_{\rm max, rms}$,
 given by Eq.~(\ref{eqn:Qmax_rms}).
 Three frequently used target nuclei:
 $\rmF$      (bottom solid        blue),
 $\rmGe$     (middle dashed       red),
 and $\rmXe$ (top    dash--dotted black)
 have been considered.
}
\label{fig:Qmax_rms-mchi}
\end{figure}

 In Sec.~\ref{sec:NR-chi-F},
 it has been found that
 the kinetic behavior
 (the recoil--angle distributions of
  the recoil direction
  and the accumulated/average recoil energies) of
 (light target nuclei like) $\rmF$
 scattered by (100-GeV) halo WIMPs
 might be approximated by
 scattering by (100-GeV) WIMPs
 moving with monotonically
 the root--mean--square velocity/average kinetic energy
 (within the 1$\sigma$ statistical uncertainties).

 On one hand,
 as shown in Fig.~\ref{fig:FQ},
 the cross section (nuclear form factor) suppression of
 (light nuclei like) $\rmF$
 is (relatively) pretty weak.
 On the other hand,
 in Fig.~\ref{fig:Qmax_rms-mchi},
 the WIMP--mass dependence of
 the maximum (prefactor) of the recoil energy $Q$
 given by Eq.~(\ref{eqn:QQ_thetaNRchi})
 with the root--mean--square velocity
 as the common velocity of incident WIMPs:
\beq
     Q_{\rm max, rms}
  =  \afrac{2 \mrN^2}{\mN} v_{\rm rms, Lab}^2
\~,
\label{eqn:Qmax_rms}
\eeq
 shows that
 the maximal recoil energy
 transferred by WIMPs
 increases basically with
 the increased mass of the target nucleus
 (and/or that of incident WIMPs).
 Hence,
 in this subsection,
 we consider
 $\rmGe$ and $\rmXe$
 as one middle--mass and one heavy target nuclei
 in our simulations,
 in order to investigate
 the target dependence of
 the recoil--angle distributions of
 the recoil direction (flux)/energy of
 scattered target nuclei.

 \def \Target      {Ge73}
 \InsertPlotNRchi
  {target&mchi=100}
  {500}
  {As in Figs.~\ref{fig:NR-F19-0100-chi-0500-\PlotNumberAa},
   except that
   a middle--mass nucleus $\rmGe$
   has been considered as our target.%
   }
 \def \Target      {Xe129}
 \InsertPlotNRchi
  {target&mchi=100}
  {500}
  {As in Figs.~\ref{fig:NR-F19-0100-chi-0500-\PlotNumberAa}
   and \ref{fig:NR-Ge73-0100-chi-0500-\PlotNumberAa},
   except that
   a heavy nucleus $\rmXe$
   has been considered as our target.%
   }

 In Figs.~\ref{fig:NR-Ge73-0100-chi-0500-\PlotNumberAa}
 and \ref{fig:NR-Xe129-0100-chi-0500-\PlotNumberAa},
 we show
 the angular distributions
 and the corresponding recoil--angle dependences of
 the recoil direction (flux)
 and the accumulated/average recoil energies of
 $\rmGe$ and $\rmXe$ target nuclei
 (scattered by 100-GeV halo WIMPs)%
\footnote{
 Interested readers can click each plot
 in Figs.~\ref{fig:NR-F19-0100-chi-0500-\PlotNumberAa},
 \ref{fig:NR-Ge73-0100-chi-0500-\PlotNumberAa},
 and \ref{fig:NR-Xe129-0100-chi-0500-\PlotNumberAa}
 to open the corresponding webpage of
 the animated demonstration
 with varying target nuclei.
}.
 Firstly,
 it can be found
 from the top--two rows of two figures
 that
 the most frequent/energetic recoil directions (angles)
 of both nuclei
 do not appear
 around the $+\zchi$--axis at all,
 but around the recoil angles of $\thetaNRchi \simeq 27.5^{\circ}$
 and $\thetaNRchi \simeq 45^{\circ}$
 for $\rmGe$
 and the recoil angles of $\thetaNRchi \simeq 20^{\circ}$
 and $\thetaNRchi \simeq 35^{\circ}$
 for $\rmXe$,
 respectively.
 This indicates clearly that,
 due to
 the (much) strong(er) cross section (nuclear form factor) suppressions
 on high recoil energies
 and thus large recoil angles,
 the heavier the mass of our target nucleus,
 the smaller the most--frequent/energetic
 recoil angles.

 Secondly,
 the right columns
 of Figs.~\ref{fig:NR-Ge73-0100-chi-0500-\PlotNumberAa}
 and \ref{fig:NR-Xe129-0100-chi-0500-\PlotNumberAa}
 show that
 the kinetic behaviors of
 both target nuclei
 (scattered by 100-GeV halo WIMPs)
 could {\em not} be approximated by
 the (dash--dotted red) average--kinetic--energy curve
 any more:
 for (middle--mass target nuclei like) $\rmGe$,
 the (black) simulated histograms
 match the average--kinetic--energy curve
 only in the small recoil angle range
 and move towards the (dash--double--dotted blue) half--average--kinetic--energy curve
 in the large recoil angle range,
 whereas,
 with the increased nuclear mass,
 for (heavy target nuclei like) $\rmXe$,
 the histograms
 move to approximately between
 the average-- and the half--average--kinetic--energy curves
 or even closer to the latter
 in the large recoil angle range.

 It is understandable that,
 once all incident WIMPs
 move (unrealistically) with
 the (higher) common root--mean--square velocity
 $v_{\rm rms, Lab}$
 (and thus the (larger) average kinetic energy $\Expv{\Echi}$),
 the large recoil angle
 (the large recoil energy) of
 the scattered target nuclei
 would be more strongly suppressed
 due to the sharply reduced cross section (nuclear form factor);
 in contrast,
 in the (unrealistic) case that
 all incident WIMPs move with
 (the lower) $v_{\rm rms, Lab} / \sqrt{2}$
 (namely only the half of the average kinetic energy),
 the suppression on
 the large recoil angle
 (but only the half of the recoil energy)
 would be somehow released.
 And in the real world,
 as demonstrated in Ref.~\cite{DMDDD-fv_eff},
 the average
 and the root--mean--square velocities of
 the 100-GeV WIMPs
 scattering off
 (heavy target nuclei like) $\rmGe$ and $\rmXe$
 would be smaller than
 those of
 the entire group of WIMPs
 impinging into the detectors.
 Thus
 the recoil--angle distributions of
 $\rmGe$ and $\rmXe$
 shift from the average--kinetic--energy curves
 to the haf--average--kinetic--energy curves.
 However,
 for cases of very heavy target nuclei
 (like%
  \OnlinePlotNR
   {NR_theta}
   {} {NR_theta}
   {\ShortFrame}
   {mchi&target=W183}
   {periodA}
   {500}
   {$\rmW$})
 scattered by very heavy (e.g.~200-GeV) WIMPs,
 due to the very strong suppression on
 high velocity/kinetic energy WIMPs,
 the incident velocity of
 most of the WIMPs
 scattering off target nuclei
 would be pretty low
 and thus
 could sometimes kick target nuclei
 with large recoil angles.

 Remind that
 the recoil energy $Q$
 appearing in the scattering validation criterion (\ref{eqn:f_NR_thetaNRchi})
 depends on the recoil angle $\thetaNRchi$
 by Eq.~(\ref{eqn:QQ_thetaNRchi}).
 Although
 the higher the incident velocity of
 the (simulated) incoming WIMP
 and/or the larger the recoil angle $\thetaNRchi$,
 the larger the recoil energy,
 which could be transferred to the scattered target nucleus,
 the key point is,
 as shown in Fig.~\ref{fig:FQ},
 the larger the recoil energy
 and/or the heavier the mass of the scattered target nucleus,
 the stronger the cross section (nuclear form factor) suppression.
 Hence,
 the heavier the mass of the target nucleus,
 the less possible that WIMPs
 with high incident velocities
 scatter off the nuclei
 with large recoil angles,
 as demonstrated
 in Figs.~\ref{fig:NR-F19-0100-chi-0500-\PlotNumberAa},
 \ref{fig:NR-Ge73-0100-chi-0500-\PlotNumberAa},
 and \ref{fig:NR-Xe129-0100-chi-0500-\PlotNumberAa}.

 Finally,
 the angular and recoil--angle distributions of
 the average recoil energies of
 two nuclei
 in the bottom rows
 of Figs.~\ref{fig:NR-Ge73-0100-chi-0500-\PlotNumberAa}
 and \ref{fig:NR-Xe129-0100-chi-0500-\PlotNumberAa}
 show that,
 although
 the maximal transferable recoil energies
 by 100-GeV WIMPs
 moving with the root--mean--square velocity
 are approximately equal
 for $\rmGe$ and $\rmXe$
 (see also Fig.~\ref{fig:Qmax_rms-mchi}),
 due to its much stronger
 cross section (nuclear form factor) suppression,
 the average recoil energy of
 the $\rmXe$ nucleus
 is clearly smaller than that of
 the $\rmGe$ nucleus.

\subsection{WIMP--mass dependence of
            the angular recoil--direction/energy distributions}
\label{sec:NR-chi-mchi}

 Following the previous Sec.~\ref{sec:NR-chi-target},
 in this subsection,
 we turn to discuss
 the WIMP--mass dependence of
 the angular distributions of
 the recoil direction/energy of
 scattered target nuclei.

 \def \Target      {F19}
 \def \WIMPmass    {0020}
 \def \EventNumber {0500}
 \InsertPlotNRchi
  {mchi&target=\Target}
  {500}
  {As in Figs.~\ref{fig:NR-F19-0100-chi-0500-\PlotNumberAa}:
   $\rmF$ has been considered as the target nucleus,
   except that
   the mass of incident WIMPs
   has been considered as light as $\mchi = 20$ GeV.%
   }
 \def \WIMPmass    {0200}
 \InsertPlotNRchi
  {mchi&target=\Target}
  {500}
  {As in Figs.~\ref{fig:NR-F19-0020-chi-0500-\PlotNumberAa}
   and \ref{fig:NR-F19-0100-chi-0500-\PlotNumberAa}:
   $\rmF$ has been considered as the target nucleus,
   except that
   the mass of incident WIMPs
   has been considered as heavy as $\mchi = 200$ GeV.%
   }

 At first,
 due to its light nuclear mass,
 the mass--dependent factor of
 the recoil energy of the $\rmF$ nucleus
 given by Eq.~(\ref{eqn:QQ_thetaNRchi}),
 $2 \mrN^2 / \mN$,
 varies from $\sim$ 10 GeV
 (for $\mchi = 20$ GeV)
 to \mbox{$\sim$ 30 GeV}
 (for $\mchi = 200$ GeV):
 only $\sim$ 3 times enlarged.
 Additionally,
 the nuclear form factor of $\rmF$
 reduces (relatively) pretty flat
 (see Fig.~\ref{fig:FQ}).
 The differences of
 the angular distributions of
 the recoil direction/energy of
 the light $\rmF$ nucleus
 scattered by light (20-GeV) WIMPs
 or by heavier (200-GeV) WIMPs
 shown in Figs.~\ref{fig:NR-F19-0020-chi-0500-\PlotNumberAa}
 and \ref{fig:NR-F19-0200-chi-0500-\PlotNumberAa}
 would be pretty small,
 but all characteristics discussed in detail below
 would already be observable.

 \def \Target      {Ge73}
 \def \WIMPmass    {0020}
 \InsertPlotNRchi
  {mchi&target=\Target}
  {500}
  {As in Figs.~\ref{fig:NR-Ge73-0100-chi-0500-\PlotNumberAa}:
   $\rmGe$ has been considered as the target nucleus,
   except that
   the mass of incident WIMPs
   has been considered as light as $\mchi = 20$ GeV.%
   }
 \def \WIMPmass    {0200}
 \InsertPlotNRchi
  {mchi&target=\Target}
  {500}
  {As in Figs.~\ref{fig:NR-Ge73-0020-chi-0500-\PlotNumberAa}
   and \ref{fig:NR-Ge73-0100-chi-0500-\PlotNumberAa}:
   $\rmGe$ has been considered as the target nucleus,
   except that
   the mass of incident WIMPs
   has been considered as heavy as $\mchi = 200$ GeV.%
   }
 \def \Target      {Xe129}
 \def \WIMPmass    {0020}
 \InsertPlotNRchi
  {mchi&target=\Target}
  {500}
  {As in Figs.~\ref{fig:NR-Xe129-0100-chi-0500-\PlotNumberAa}:
   $\rmXe$ has been considered as the target nucleus,
   except that
   the mass of incident WIMPs
   has been considered as light as $\mchi = 20$ GeV.%
   }
 \def \WIMPmass    {0200}
 \InsertPlotNRchi
  {mchi&target=\Target}
  {500}
  {As in Figs.~\ref{fig:NR-Xe129-0020-chi-0500-\PlotNumberAa}
   and \ref{fig:NR-Xe129-0100-chi-0500-\PlotNumberAa}:
   $\rmXe$ has been considered as the target nucleus,
   except that
   the mass of incident WIMPs
   has been considered as heavy as $\mchi = 200$ GeV.%
   }

 In contrast,
 due to their much larger enlargements of
 the mass--dependent factor of
 the recoil energy:
 varies from $\sim$ 7 GeV
 to $\sim$ 76 GeV
 ($\sim$ 11 times enlarged)
 for the $\rmGe$ target
 and
 varies from $\sim$ 5 GeV
 to $\sim$ 94 GeV
 ($\sim$ 19 times enlarged)
 for the $\rmXe$ target,
 as well as
 their much more sharply reduced cross section (nuclear form factor)
 with the increased recoil energy,
 the angular recoil--direction/energy distributions of
 the middle--mass $\rmGe$ nucleus
 and,
 in particular,
 the heavy $\rmXe$ nucleus
 scattered by 20-GeV
 and 200-GeV WIMPs
 presented from Figs.~\ref{fig:NR-Ge73-0020-chi-0500-\PlotNumberAa}
 to \ref{fig:NR-Xe129-0200-chi-0500-\PlotNumberAa}
 show a clear WIMP--mass dependence%
\footnote{
 Interested readers can click each plot
 in Figs.~\ref{fig:NR-F19-0020-chi-0500-\PlotNumberAa}
 to \ref{fig:NR-Xe129-0200-chi-0500-\PlotNumberAa}
 to open the corresponding webpage of
 the animated demonstration
 with the varying WIMP mass
 (for more considered target nuclei).
}:
 for a given target nucleus,
 the larger the WIMP mass,
 the smaller the most frequent/energetic recoil angles;
 once the WIMP mass is
 as heavy as $\cal O$(200 GeV),
 only a small fraction of incident halo WIMPs
 (with low incident velocities)
 could scatter off (heavy target nuclei like) $\rmXe$
 with large recoil angles.

 Moreover,
 the flat
 recoil--angle distributions of
 the average recoil energies of
 the scattered $\rmGe$ and $\rmXe$ nuclei
 in the high--$\thetaNRchi$ ranges
 shown in the bottom--right frames of
 Figs.~\ref{fig:NR-Ge73-0200-chi-0500-\PlotNumberAa}
 and \ref{fig:NR-Xe129-0200-chi-0500-\PlotNumberAa}
 would confirm our observation
 discussed in Sec.~\ref{sec:NR-chi-target}:
 although
 the WIMP kinetic energy
 and thus the maximal theoretically transferable recoil energy
 (with a given incident velocity)
 increases rapidly
 with the increase of the considered WIMP mass
 (see Fig.~\ref{fig:Qmax_rms-mchi}),
 the actual increase of
 the average recoil energy
 (in particular,
  in the high--$\thetaNRchi$ range)
 would strongly be limited,
 due to,
 as already emphasized,
 the cross section (nuclear form factor) suppression.

\subsection{Annual modulation of
            the angular recoil--direction/energy distributions}
\label{sec:NR-chi-annual}

 Due to the orbital rotation of the Earth around the Sun,
 the relative velocity of incident halo WIMPs
 with respect to our laboratory/detector
 varies annually.
 Then
 the flux and the kinetic energy of incident WIMPs
 as well as
 the recoil energy of target nuclei
 scattered by incident WIMPs
 should in turn vary annually.
 Hence,
 in this subsection,
 we discuss briefly
 the annual modulation of
 the angular ($\thetaNRchi$) distributions of
 the recoil direction (flux)
 and the accumulated/average recoil energies of
 WIMP--scattered target nuclei
 in the incoming--WIMP coordinate system.

 \def \Target        {F19}
 \def \WIMPmass      {0100}
 \def \Perioda       {\PeriodCa}
 \def \Periodb       {\PeriodCb}
 \def \Periodc       {\PeriodCc}
 \def \Periodd       {\PeriodCd}
 \def \PlotNumbera   {\PlotNumberCa}
 \def \PlotNumberb   {\PlotNumberCb}
 \def \PlotNumberc   {\PlotNumberCc}
 \def \PlotNumberd   {\PlotNumberCd}
 \InsertPlotNRAnnual
  {theta}
  {As in the right column
   of Figs.~\ref{fig:NR-F19-0100-chi-0500-\PlotNumberAa}:
   $\rmF$ target nuclei
   scattered by 100-GeV WIMPs,
   except that
   500 accepted events on average
   in each 60-day observation period
   of four {\em advanced} seasons
   \cite{DMDDD-N, DMDDD-3D-WIMP-N}
   have been considered.%
   }
 \def \Target        {Ge73}
 \InsertPlotNRAnnual
  {theta}
  {As in Figs.~\ref{fig:NR_theta-F19-0100-chi-0500-\PlotNumberCa},
   except that
   a middle--mass nucleus $\rmGe$
   has been considered as our target.%
   }
 \def \Target        {Xe129}
 \InsertPlotNRAnnual
  {theta}
  {As in Figs.~\ref{fig:NR_theta-F19-0100-chi-0500-\PlotNumberCa}
   and \ref{fig:NR_theta-Ge73-0100-chi-0500-\PlotNumberCa},
   except that
   a heavy nucleus $\rmXe$
   has been considered as our target.%
   }

 In Figs.~\ref{fig:NR_theta-F19-0100-chi-0500-\PlotNumberCa},
 \ref{fig:NR_theta-Ge73-0100-chi-0500-\PlotNumberCa},
 and \ref{fig:NR_theta-Xe129-0100-chi-0500-\PlotNumberCa},
 we show
 the recoil--angle distributions of
 the event number (top),
 the accumulated (middle)
 and the average (bottom) recoil energies of
 three target nuclei
 scattered by 100-GeV WIMPs
 with
 500 accepted events on average
 in each 60-day observation period
 of four {\em advanced} seasons
 \cite{DMDDD-N}%
\footnote{
 Interested readers can click each row
 in Figs.~\ref{fig:NR_theta-F19-0100-chi-0500-\PlotNumberCa},
 \ref{fig:NR_theta-Ge73-0100-chi-0500-\PlotNumberCa},
 and \ref{fig:NR_theta-Xe129-0100-chi-0500-\PlotNumberCa}
 to open the webpage of
 the animated demonstration
 for the corresponding annual modulation
 (and for more considered WIMP masses
  and target nuclei).
}.

 Although,
 considering the large (1$\sigma$) statistical uncertainties,
 it should practically be very difficult
 to identify these tiny annual variations
 with $\cal O$(500) or even a few thousands of
 WIMP scattering events
 observed in each season
 (in several consecutive years),
 it would still be worth to note here that
 there would indeed be
 tiny differences between
 the distributions of
 the recoil angle
 as well as
 those of
 the accumulated/average recoil energies
 in different (advanced) seasons:
 while more scattering WIMPs
 would have higher (lower) incident velocity
 in the advanced summer (winter)
 \cite{DMDDD-fv_eff},
 firstly,
 more scattered target nuclei
 would go to smaller (larger) recoil angles
 in summer (winter);
 secondly,
 the accumulated recoil energy
 in the small (or even all) recoil angles
 would be maximal (minimal)
 in summer (winter);
 and,
 finally,
 the average recoil energy (per event)
 in all recoil angles
 would also be maximal (minimal)
 in summer (winter).

%% file: sec-NR_ang-Lab.tex
%
%
\section{Angular distributions of
         the recoil direction/energy
         in the laboratory frame}
\label{sec:NR_ang-Lab}

 In this section,
 we move to discuss
 the angular distributions of
 the recoil direction (flux)
 as well as
 the accumulated and the average recoil energies of
 scattered target nuclei
 observed in the laboratory coordinate system.
 Note that,
 instead of
 the angular recoil--direction/energy distributions
 in the incoming--WIMP coordinate system
 demonstrated in Sec.~\ref{sec:NR-chi},
 this 3-D information on
 elastic WIMP--nucleus scattering signals
 could practically be measured
 in directional direct detection experiments.

 \def \Target         {F19}
 \def \ShortFrame     {Lab}
 \def \EventNumber    {0500}
 \def \PlotNumber     {\PlotNumberAa}
 \def \LabName        {ANDES}
 \def \LabLocation    {(30.19$^{\circ}$S, 69.82$^{\circ}$W)}
 \InsertPlotNRLabmchi
  {The angular distributions of
   the recoil direction (flux) (top),
   the accumulated (middle)
   and the average (bottom) recoil energies
   of $\rmF$
   observed in the laboratory coordinate system of
   the \LabName\ laboratory
   \LabLocation\
   \cite{\LabName}
   in unit of the all--sky average values.
   The masses of incident WIMPs
   have been set as $\mchi = 20$ GeV (left),
   100 GeV (center),
   and 200 GeV (right),
   respectively.
   Other simulation setup and notations are the same
   as in Figs.~\ref{fig:NR-F19-0020-chi-0500-\PlotNumberAa},
   \ref{fig:NR-F19-0100-chi-0500-\PlotNumberAa},
   and \ref{fig:NR-F19-0200-chi-0500-\PlotNumberAa}.%
   }
 \def \Target         {Ge73}
 \InsertPlotNRLabmchi
  {As in Figs.~\ref{fig:NR-F19-Lab-0500-\PlotNumberAa-\LabName},
   except that
   a middle--mass nucleus $\rmGe$
   has been considered as our target.%
   }
 \def \Target         {Xe129}
 \InsertPlotNRLabmchi
  {As in Figs.~\ref{fig:NR-F19-Lab-0500-\PlotNumberAa-\LabName}
   and \ref{fig:NR-Ge73-Lab-0500-\PlotNumberAa-\LabName},
   except that
   a heavy nucleus $\rmXe$
   has been considered as our target.%
   }

 In Figs.~\ref{fig:NR-F19-Lab-0500-\PlotNumberAa-\LabName},
 \ref{fig:NR-Ge73-Lab-0500-\PlotNumberAa-\LabName},
 and \ref{fig:NR-Xe129-Lab-0500-\PlotNumberAa-\LabName},
 we show
 the angular distributions of
 the recoil direction (flux) (top),
 the accumulated (middle)
 and the average (bottom) recoil energies of
 the light $\rmF$,
 the middle--mass $\rmGe$,
 and the heavy $\rmXe$ nuclei
 observed in the laboratory coordinate system of
 the \LabName\ laboratory
 \LabLocation,
 which would be
 the second functionable underground laboratory
 in the Southern Hemisphere
 \cite{\LabName},
 in unit of the all--sky average values,
 respectively.
 As in Sec.~\ref{sec:NR-chi},
 three different masses of
 incident halo WIMPs:
  20 GeV (left),
 100 GeV (center),
 and 200 GeV (right),
 have been considered%
\footnote{
 Interested readers can click each row
 of Figs.~\ref{fig:NR-F19-Lab-0500-\PlotNumberAa-\LabName},
 \ref{fig:NR-Ge73-Lab-0500-\PlotNumberAa-\LabName},
 and \ref{fig:NR-Xe129-Lab-0500-\PlotNumberAa-\LabName}
 to open the corresponding webpage of
 the animated demonstration
 with the varying WIMP mass
 (for more considered target nuclei
  as well as
  other underground laboratories).
 One can also check ``Target dependence''
 for the ``animation mode''
 on the webpage
 to see the animated demonstration
 with varying target nuclei.
}.

 First of all,
 for all three considered target nuclei
 and three simulated WIMP masses%
\footnote{
 Currently,
 we consider only rather heavy WIMP masses of $\mchi \ge 20$ GeV.
 For light WIMP masses of \mbox{$\mchi \le 20$ GeV},
 more detailed investigations
 are needed and
 will be announced later.
},
 it can be seen directly that
 the basic distribution patterns of
 the most frequent/energetic
 recoil directions
 (the top and the middle rows)
 spread approximately symmetrically
 with respect to the azimuthal angle of 180$^{\circ}$.

 Moreover,
 similar to
 the angular recoil--direction/energy distributions of
 three considered target nuclei
 in the incoming--WIMP coordinate system
 discussed previously,
 one can find that,
 firstly,
 for one given target nucleus,
 the heavier its mass,
 the more obvious the (qualitative) differences
 between the angular distribution patterns
 simulated with different WIMP masses.
 This would imply that,
 qualitatively speaking,
 the 3-D recoil distribution patterns of
 heavy target nuclei
 (e.g.~$\rmXe$)
 would be more useful
 for identifying the mass (range) of halo WIMPs.
 Nevertheless,
 as shown quantitatively
 in the right columns
 of Figs.~\ref{fig:NR-F19-0020-chi-0500-\PlotNumberAa}
 to \ref{fig:NR-Xe129-0200-chi-0500-\PlotNumberAa},
 the all--sky average values of
 the accumulated/average recoil energies
 depend strongly on the simulated WIMP mass
 as well as
 on the mass of the target nucleus.
 This means that
 the normalization standards of
 the distribution patterns
 shown in each (corresponding) row
 of Figs.~\ref{fig:NR-F19-Lab-0500-\PlotNumberAa-\LabName},
 \ref{fig:NR-Ge73-Lab-0500-\PlotNumberAa-\LabName},
 and \ref{fig:NR-Xe129-Lab-0500-\PlotNumberAa-\LabName}
 are different.
 Hence,
 the measured 3-D information about
 the angular distributions of
 the recoil direction (flux)/energy
 of light nuclei
 (e.g.~$\rmF$)
 would also be helpful
 for constraining the mass (range) of halo WIMPs.

 On the other hand,
 it could also be clearly observed that,
 for a given WIMP mass,
 the larger the mass difference
 between two target nuclei
 (and the larger the considered WIMP mass),
 the larger
 the differences
 between their angular distribution patterns of
 the recoil direction/energy.
 This indicates that,
 by comparing/combining
 the angular recoil--direction/energy distributions of
 different target nuclei,
 we could pin down
 the mass (range) of halo WIMPs
 (more precisely).

%% file: sec-NR_ang-Eq.tex
%
%
\section{Angular distributions of
         the recoil direction/energy
         in the Equatorial frame}
\label{sec:NR_ang-Eq}

 Finally,
 we present
 in this section
 the angular distributions of
 the recoil direction (flux)
 as well as
 the accumulated and the average recoil energies of
 scattered target nuclei
 observed in the Equatorial coordinate system.
 Note that,
 by definition,
 the Equatorial coordinate system
 is laboratory/location independent
 \cite{DMDDD-N}.
 Hence,
 considering the very low theoretically estimated event rate,
 combining and analyzing WIMP scattering events
 off the same target nucleus
 provided however from different underground laboratories
 in the Equatorial coordinate system
 would be practically a very useful strategy.

 \def \Target      {F19}
 \def \ShortFrame  {Eq}
 \def \EventNumber {0500}
 \def \PlotNumber  {\PlotNumberAa}
 \InsertPlotNRmchi
  {The angular distributions of
   the recoil direction (flux) (top),
   the accumulated (middle)
   and the average (bottom) recoil energies
   of $\rmF$
   observed in the Equatorial coordinate system
   in unit of the all--sky average values.
   The masses of incident WIMPs
   have been set as $\mchi = 20$ GeV (left),
   100 GeV (center),
   and 200 GeV (right),
   respectively.
   The dark--green star indicates
   the theoretical main direction of incident WIMPs
   in the Equatorial coordinate system
   \cite{Bandyopadhyay10}:
   42.00$^{\circ}$S, 50.70$^{\circ}$W.
   The simulation setup and notations are the same
   as in Figs.~\ref{fig:NR-F19-Lab-0500-\PlotNumberAa-ANDES}.%
   }
 \def \Target      {Ge73}
 \InsertPlotNRmchi
  {As in Figs.~\ref{fig:NR-F19-Eq-0500-\PlotNumberAa},
   except that
   a middle--mass nucleus $\rmGe$
   has been considered as our target.%
   }
 \def \Target      {Xe129}
 \InsertPlotNRmchi
  {As in Figs.~\ref{fig:NR-F19-Eq-0500-\PlotNumberAa}
   and \ref{fig:NR-Ge73-Eq-0500-\PlotNumberAa},
   except that
   a heavy nucleus $\rmXe$
   has been considered as our target.%
   }

 In Figs.~\ref{fig:NR-F19-Eq-0500-\PlotNumberAa},
 \ref{fig:NR-Ge73-Eq-0500-\PlotNumberAa},
 and \ref{fig:NR-Xe129-Eq-0500-\PlotNumberAa},
 we show
 the angular distributions of
 the recoil direction (flux) (top),
 the accumulated (middle)
 and the average (bottom) recoil energies of
 the light $\rmF$,
 the middle--mass $\rmGe$,
 and the heavy $\rmXe$ nuclei
 observed in the Equatorial coordinate system
 in unit of the all--sky average values.
 As in Secs.~\ref{sec:NR-chi} and \ref{sec:NR_ang-Lab},
 three different masses of
 incident halo WIMPs:
  20 GeV (left),
 100 GeV (center),
 and 200 GeV (right),
 have been considered%
\footnote{
 Interested readers can click each row
 in Figs.~\ref{fig:NR-F19-Eq-0500-\PlotNumberAa},
 \ref{fig:NR-Ge73-Eq-0500-\PlotNumberAa},
 and \ref{fig:NR-Xe129-Eq-0500-\PlotNumberAa}
 to open the corresponding webpage of
 the animated demonstration
 with the varying WIMP mass
 (for more considered target nuclei).
 One can also check ``Target dependence''
 for the ``animation mode''
 on the webpage
 to see the animated demonstration
 with varying target nuclei.
}.
 The dark--green star in each plot indicates
 the theoretical main direction of incident WIMPs
 in the Equatorial coordinate system:
 \cite{Bandyopadhyay10}:
 42.00$^{\circ}$S, 50.70$^{\circ}$W.

 At first,
 similar to
 the angular distributions of
 the recoil direction/energy of
 three considered target nuclei
 in the laboratory coordinate system
 presented in Sec.~\ref{sec:NR_ang-Lab},
 one can find that,
 the heavier the mass of one given target nucleus
 or the larger the mass difference
 between two target nuclei
 (and the larger the simulated WIMP mass),
 the more obvious the pattern differences
 between the angular recoil--direction/energy distributions of
 the scattered target nucleus
 with different WIMP masses
 or between those of different target nuclei.
 Hence,
 by combining and/or comparing
 the (quantitative) nuclear recoil flux and energy
 or even only their (qualitative) distribution patterns
 with the same/different target nuclei
 (in different underground laboratories)
 in the Equatorial coordinate system,
 one could constrain/identify the mass (range) of halo WIMPs.

 More importantly,
 the angular distributions of
 the recoil direction/energy of
 all three target nuclei
 with three considered WIMP masses
 show clearly the anisotropies.
 However,
 unexpectedly,
 neither
 the centers of the most frequent recoil directions
 nor
 those of the most energetic ones of
 all simulated target nucleus--WIMP mass combinations
 (the top and the middle rows)
 could match
 the theoretical main direction of
 the WIMP wind
 (the dark--green star).
 Comparing with
 the small (northwest) deviations of
 the most frequent and the most energetic incident directions of
 halo WIMPs
 presented in Refs.~\cite{DMDDD-N, DMDDD-P},
 the northerly shifts of
 (the centers of)
 the most frequent/energetic recoil directions
 seem to be much more obvious,
 and these shifts increase
 with the increased nuclear and/or WIMP mass.
 Interestingly,
 the hot--points of
 the angular distributions of
 the ``average'' recoil energy
 (the bottom rows) of
 all simulated target nucleus--WIMP mass combinations
 seem to be approximately centered
 around the theoretical main WIMP direction,
 with nevertheless
 some small (but observable)
 target/WIMP--mass dependent pattern differences.

\subsection{Annual modulation of
            the angular recoil--direction/energy distributions}
\label{sec:NR_ang-Eq-annual}

 The original purpose of
 directional direct Dark Matter detection experiments
 is to observe the ``diurnal'' modulated anisotropy
 (the directionality) of
 the recoil direction of
 elastic WIMP--nucleus scattering events
 \cite{Ahlen09, Mayet16}.
 On the other hand,
 in Refs.~\cite{DMDDD-N, DMDDD-P},
 we demonstrated that,
 considering the very low theoretically estimated event rate,
 it would be equivalent or even better
 to observe the ``annual'' modulated anisotropy of
 the recoil direction of
 WIMP--scattered target nuclei.
 Hence,
 in this subsection,
 we discuss
 the annual modulation of
 the angular distributions of
 the recoil direction/energy
 observed in the Equatorial coordinate system.

 \def \Target        {F19}
 \def \WIMPmass      {0100}
 \def \ShortFrame    {Eq}
 \def \Perioda       {\PeriodCa}
 \def \Periodb       {\PeriodCb}
 \def \Periodc       {\PeriodCc}
 \def \Periodd       {\PeriodCd}
 \def \PlotNumbera   {\PlotNumberCa}
 \def \PlotNumberb   {\PlotNumberCb}
 \def \PlotNumberc   {\PlotNumberCc}
 \def \PlotNumberd   {\PlotNumberCd}
 \InsertPlotNRAnnual
  {ang}
  {As in Figs.~\ref{fig:NR-F19-Eq-0500-\PlotNumberAa}(b):
   $\rmF$ target nuclei
   scattered by 100-GeV WIMPs,
   except that
   500 accepted events on average
   in each 60-day observation period
   of four advanced seasons
   \cite{DMDDD-N, DMDDD-3D-WIMP-N}
   have been considered.
   Besides the dark--green star
   indicating
   the theoretical main direction of
   the WIMP wind,
   the blue--yellow point in each plot
   indicates
   the opposite direction of
   the Earth's velocity relative to the Dark Matter halo
   on the central date of the observation period
   \cite{DMDDD-N}.%
   }
 \def \Target        {Ge73}
 \InsertPlotNRAnnual
  {ang}
  {As in Figs.~\ref{fig:NR_ang-F19-0100-Eq-0500-\PlotNumberCa},
   except that
   a middle--mass nucleus $\rmGe$
   has been considered as our target.%
   }
 \def \Target        {Xe129}
 \InsertPlotNRAnnual
  {ang}
  {As in Figs.~\ref{fig:NR_ang-F19-0100-Eq-0500-\PlotNumberCa}
   and \ref{fig:NR_ang-Ge73-0100-Eq-0500-\PlotNumberCa},
   except that
   a heavy nucleus $\rmXe$
   has been considered as our target.%
   }

 In Figs.~\ref{fig:NR_ang-F19-0100-Eq-0500-\PlotNumberCa},
 \ref{fig:NR_ang-Ge73-0100-Eq-0500-\PlotNumberCa},
 and \ref{fig:NR_ang-Xe129-0100-Eq-0500-\PlotNumberCa},
 we show
 the angular distributions of
 the recoil direction (flux) (top),
 the accumulated (middle)
 and the average (bottom) recoil energies of
 the light $\rmF$,
 the middle--mass $\rmGe$,
 and the heavy $\rmXe$ nuclei
 scattered by 100-GeV WIMPs
 observed in the Equatorial coordinate system
 in unit of the all--sky average values
 with
 500 accepted events on average
 in each 60-day observation period
 of four advanced seasons
 \cite{DMDDD-N}%
\footnote{
 Interested readers can click each row
 in Figs.~\ref{fig:NR_ang-F19-0100-Eq-0500-\PlotNumberCa}
 to \ref{fig:NR_ang-Xe129-0200-Eq-0500-\PlotNumberCa}
 to open the webpage of
 the animated demonstration
 for the corresponding annual modulation
 (and for more considered WIMP masses
  and target nuclei).
}.
 Besides the dark--green star
 indicating
 the theoretical main direction of
 the WIMP wind,
 the blue--yellow point in each plot
 indicates
 the opposite direction of
 the Earth's velocity relative to the Dark Matter halo
 on the central date of the observation period
 \cite{DMDDD-N}.

 It can be found that,
 firstly,
 while
 the angular distribution patterns of
 the ``average'' recoil energy of
 three considered target nuclei
 (the bottom rows)
 show indeed approximately circular clockwise rotations
 following the instantaneous theoretical main direction of
 the incident halo WIMPs,
 the approximate rotation centers of
 the distribution patterns of
 the recoil direction and
 the accumulated recoil energy
 (the top and the middle rows)
 would,
 as observed previously,
 clearly shift northwesterly.

 Moreover,
 the annual variations of
 the angular recoil--direction distributions of
 (middle--mass nuclei like) $\rmGe$
 and (heavy nuclei like) $\rmXe$
 seem {\em not} to be circular,
 but oscillated somehow {\em latitudinally}.
 And the shapes of the most frequent/energetic recoil directions
 seem also to be somehow {\em longitudinally} squeezed;
 the heavier the nuclear (and/or the WIMP) mass,
 the stronger the pattern distortion would be.
 So far
 we have no reasonable explanation(s)
 for this (unexpected) observations.

 \def \Target        {Xe129}
 \def \WIMPmass      {0200}
 \InsertPlotNRAnnual
  {ang}
  {As in Figs.~\ref{fig:NR_ang-Xe129-0100-Eq-0500-\PlotNumberCa}:
   $\rmXe$ has been considered as the target nucleus,
   except that
   the mass of incident WIMPs
   has been considered as heavy as $\mchi = 200$ GeV.%
   }

 In Figs.~\ref{fig:NR_ang-Xe129-0200-Eq-0500-\PlotNumberCa},
 we show
 the angular distributions of
 the recoil direction/energy of
 the heavy $\rmXe$ nuclei
 with a raised WIMP mass of $\mchi = 200$ GeV,
 as a demonstration of
 the WIMP--mass dependence of
 the annual modulation of
 the angular recoil--direction/energy distributions.
 The northwesterly shifts of
 the approximate rotation centers of
 the distribution patterns of
 the recoil direction and
 the accumulated recoil energy
 as well as
 the longitudinal squeeze of
 the most frequent/energetic recoil directions
 become now more obvious.

%% file: sec-references.tex
%
%

%
%